%% file: piki_user_interface.tex
\newcommand{\arXiv}{}
\ifdefined\arXiv
\documentclass{article}
\pdfoutput=1
\usepackage[margin=.8in]{geometry}
\usepackage{authblk}

\usepackage{graphicx}
\usepackage{natbib}
\else
\documentclass[manuscript]{acmart}

\AtBeginDocument{%
  }

\copyrightyear{2023} 
\acmYear{2023} 
\setcopyright{acmlicensed}\acmConference[RecSys '23]{Seventeenth ACM Conference on Recommender Systems}{September 18--22, 2023}{Singapore, Singapore}
\acmBooktitle{Seventeenth ACM Conference on Recommender Systems (RecSys '23), September 18--22, 2023, Singapore, Singapore}
\acmPrice{15.00}
\acmDOI{10.1145/3604915.3608846}
\acmISBN{979-8-4007-0241-9/23/09}


\fi

\usepackage{amsmath}
\usepackage{amsfonts}
\usepackage{balance}  

\usepackage{tabularx}

\usepackage{hyperref}
\usepackage{cleveref}

\hypersetup{colorlinks = true,
            linkcolor = blue,
            urlcolor  = blue,
            citecolor = blue,
            anchorcolor = blue}

\usepackage{algpseudocode}
\usepackage{bm}
\usepackage{comment}
\usepackage{array}
\ifdefined\arXiv
\else
\floatstyle{plaintop}
\restylefloat{table}
\fi
\ifdefined\VLDB
\newcommand{\subparagraph}{}
\fi
\let\proof\relax
\let\endproof\relax
\usepackage{amsthm}
\usepackage{amsmath}
\usepackage{algorithm}
\usepackage{subfig}
\usepackage{titlesec}
\usepackage{blindtext}
\usepackage{multirow}
\usepackage{yfonts}
\usepackage{xcolor}
\usepackage{url}
\usepackage{balance}
\usepackage{nopageno}
\usepackage{enumitem}

\usepackage[subtle]{savetrees}

\usepackage[group-separator={,}]{siunitx}
\usepackage[tableposition=top]{caption}
\usepackage[labelfont=bf]{caption}

\ifdefined \arXiv
\usepackage{natbib}
\usepackage{booktabs}
\usepackage{placeins}

\fi

\newcommand*{\inlineequation}[2][]{%
	\begingroup
	\refstepcounter{equation}%
	\ifx\\#1\\%
	\else
	\label{#1}%
	\fi
	\relpenalty=10000 %
	\binoppenalty=10000 %
	\ensuremath{%
		#2%
	}%
	~\@eqnnum
	\endgroup
}
\algtext*{EndFunction}
\algtext*{EndIf}
\algtext*{EndFor}
\algtext*{EndWhile}
\algtext*{EndIf}

\newif\ifcomm
\commtrue
\ifcomm
\else
\commfalse
\fi
\ifcomm
	\newcommand{\mycomm}[3]{{\footnotesize{{\color{#2} \textbf{[#1: #3]}}}}}
	\newcommand{\CRdel}[1]{\textcolor{red}{\sout{#1}}}
\else
    \newcommand{\mycomm}[3]{}
    \newcommand{\CRdel}[1]{}
\fi

\begin{document}


\title{Interface Design to Mitigate Inflation in Recommender Systems}

\ifdefined\arXiv

\author{Rana Shahout \qquad Yehonatan Peisakhovsky \qquad Sasha Stoikov \qquad Nikhil Garg}
\date{}

\else
\author{Rana Shahout}
\affiliation{%
  \institution{Harvard University}
  \city{Cambridge}
  \country{USA}
}
\author{Yehonatan Peisakhovsky}
\affiliation{%
  \institution{Technion}
  \city{Haifa}
  \country{Israel}
}
\author{Sasha Stoikov}
\affiliation{%
  \institution{Cornell Tech}
  \city{New York}
  \country{USA}
}
\author{Nikhil Garg}
\affiliation{%
  \institution{Cornell Tech}
    \city{New York}
  \country{USA}}
\fi

\renewcommand{\epsilon}{\varepsilon}
\ifdefined\arXiv
\maketitle
\fi

\input{abstract}

\maketitle

\input{introduction}
\input{related_work}

\input{platform}

\input{user_heterogeinety_interface}
\input{simulation_framework}
\input{experiment}
\input{discussion}

\newpage
\ifdefined \arvix
\bibliographystyle{plainnat} 
\else
\bibliographystyle{ACM-Reference-Format}
\fi
\bibliography{refs}

\ifdefined\arXiv
\FloatBarrier
\pagebreak
\appendix
\input{appendix}

\fi
\end{document}

%% file: abstract.tex
\begin{abstract}

Recommendation systems rely on user-provided data to learn about item quality and provide personalized recommendations. An implicit assumption when aggregating ratings into item quality is that ratings are strong indicators of item quality. In this work, we test this assumption using data collected from a music discovery application. Our study focuses on two factors that cause rating inflation: heterogeneous user rating behavior and the dynamics of personalized recommendations.
We show that user rating behavior substantially varies by user, leading to item quality estimates that reflect the users who rated an item more than the item quality itself. Additionally, items that are more likely to be shown via personalized recommendations can experience a substantial increase in their exposure and potential bias toward them. To mitigate these effects, we analyze the results of a randomized controlled trial in which the rating interface was modified. The test resulted in a substantial improvement in user rating behavior and a reduction in item quality inflation. These findings highlight the importance of carefully considering the assumptions underlying recommendation systems and designing interfaces that encourage accurate rating behavior.

\end{abstract}

%% file: introduction.tex
\section{Introduction}

\label{sec:intro}

Recommendation and rating systems have emerged as crucial components of numerous online platforms, ranging from e-commerce to social media. These systems depend on user data, including both {implicit} indicators like browsing behavior and {explicit} such as rating scores. This data provides insights into the overall quality\footnote{Taste in many items, including music and movies, is inherently subjective. However, many platforms (such as Spotify, Rotten Tomatoes, and IMDb) calculate overall popularity scores or ratings -- alongside attempting to measure heterogeneous taste} of items and heterogeneous preferences, and ultimately provide personalized guidance to users.

Extensive research has been conducted on how to estimate item quality and user-item preferences and make personalized recommendations based on user data. Various approaches have been proposed, such as computing the sample mean of ratings for each item or leveraging matrix factorization or deep learning methods to generate recommendations.
In contrast, this paper focuses on the data collection process that provides input to these methods. 

Specifically, we explore how the rating data collection process affects item quality estimation and recommendation accuracy.
Here, we focus on two aspects of the data collection process: (i) user rating inflation and heterogeneity; and (ii) algorithmic bias caused by personalized recommendations.

Although previous studies have touched upon these issues, our analysis reveals that the resulting \textit{bias} and \textit{variance} in estimating item quality due to these effects are substantial. We further analyze a randomized controlled trial aimed at mitigating user rating inflation and the resulting heterogeneity in ratings. In particular, we tackle the following challenges:

\textbf{First}, we show that there is substantial user-level heterogeneity in what ratings mean (heterogeneous measurement). For example, picky users give lower ratings on average than generous users, even for similar item qualities or experiences. Thus, with naive methods, \textit{estimated overall item qualities and rankings may say more about the raters themselves than about the items being rated}. This effect is more pronounced and differentially affects items due to {cold-start-induced \textit{variance}} when new items have fewer ratings and thus are more susceptible to the idiosyncratic behavior of individual users. Depending on the platform, such heterogeneity may have substantial downstream economic consequences for item producers.

\textbf{Second}, the rating interface is a critical platform design choice that can impact recommendation systems' accuracy and effectiveness. For instance, recent research~\cite{garg2021designing} has demonstrated that changing the rating options from numerical to positive-skewed adjectives on an online labor market can result in more informative ratings. However, the extent to which interface design affects users' heterogeneous behavior and, subsequently, the impact on recommendation remains an open question. It is possible that certain interface designs produce more homogeneous user behavior, resulting in more informative item quality estimates.

To examine these effects, we analyze a natural experiment and a randomized controlled trial (RCT) on a music discovery platform. The initial platform change altered the user interface to introduce wait times before users could select their rating options (``dislike'', ``like'' or ``super-like''). The RCT further altered how long users had to wait before options were unlocked.
We find that introducing timers substantially changed user rating behavior, inducing more ``pickiness'' than before and reducing across-user behavior variance. We further find evidence that this reduction in user behavior heterogeneity improved the system's effectiveness. While a user's rating for a given item was \textit{more a function of the user who was shown the item} than the item quality (as estimated by other ratings) both before and after the change, the change increased the relative importance of item quality. 


\textbf{Third}, we investigate the relationship between personalized recommendations and user rating behavior. Using the Piki dataset\footnote{The full (user-anonymized) data is available online: \url{https://github.com/sstoikov/piki-music-dataset}.}, we conducted exploratory analyses to understand how item quality estimates are related to how often that item is recommended via personalized recommendation. Our findings reveal that randomness in personalization can significantly influence item quality estimates. This is beyond what can be expected from the fact that higher-quality items are more likely to be recommended.

	%

Overall, our work has several implications for the design of rating and recommendation systems as well as future research in this area. Specifically, our findings demonstrate that user behavior heterogeneity can significantly impact a system's ability to learn item quality and provide effective recommendations.
Our results indicate that item quality estimates reflect not only item quality but also the users who rated the items and the method used to select those users. While our work highlights interface design as a potential solution, algorithmic design, such as correcting for personalization and user heterogeneity, may also play a crucial role in improving the accuracy and fairness of item quality estimation and recommendation systems.

%% file: related_work.tex
\subsection{Related Work}
\label{sec:related_works}
This paper sits at the intersection of the literature of several communities: rating inflation, interface design and recommendation systems, and the economic implications thereof. 

\textit{Ratings inflation.}  Ratings are examples of \textit{measurements}, i.e., assignments of numbers to the construct (`how much a user enjoyed an item') that we aim to measure (see, e.g., \citet{allen2001introduction} for an overview). The literature discusses many potential reasons that measurement may not be an accurate reflection of the underlying construct, including errors in the measurement process, the use of imprecise or inadequate measurement tools, and the influence of external factors on the measurement (see \citet{bound2001measurement} for a review on biases due to measurement error in survey data).

The rating literature primarily considers \textit{inflation} error: empirically, ratings on marketplaces tend to be high, with most items (sellers, producers, freelancers) receiving positive ratings the vast majority of the time (see, e.g., \citet{filippas2018reputation}). Such inflation has substantial downstream economic consequences, in particular, that high-quality items are difficult to distinguish from lower-quality ones. 
\citet{Aziz2020TheCO}
conclude from a quasi-experiment on a restaurant ratings platform, that rating inflation leads to less exploration of new restaurants and a greater concentration of sales to more popular restaurants. To add to this literature, our work documents \textit{heterogeneous} ratings inflation and partially addresses it via interface design.




\textit{Data interface designs.} There is similarly a large literature on the design of data collection processes to minimize measurement error (see e.g., \citet{krosnick2018questionnaire} for a survey on questionnaire design). Given the importance of implicit feedback in computational systems \cite{claypool2001implicit,hu2008collaborative,jiang2015automatic,radlinski2008does,hongyiwen}, there has been substantial work on understanding \cite{kim2014modeling, liu2010understanding,huang2012user} and \textit{shaping} \cite{schnabel2019shaping} implicit feedback through interface design. Similarly, user interfaces shape how users provide and perceive explicit ratings and recommendations \cite{pu2008user,adomavicius2011recommender,cosley2003seeing,felfernig2007knowledge}. Most related is the work in rating system interface design, especially to counter inflation in marketplaces. \citet{garg2021designing} report the result of a randomized controlled trial in which the rating interface was changed. The study finds that the positive-skewed verbal rating scales lead to rating distributions that significantly reduce rating inflation and are more informative about seller quality than numeric ratings. Similarly, \citet{garg2019designing} derive the \textit{optimal} rating distribution (joint distribution between seller quality and probability of positive rating in a binary setting), when the goal is to accurately learn about item quality. Other solutions to rating inflation include modifying incentives or behavioral effects that are conjectured to cause inflation \cite{fradkin2015bias}. This paper adds to this literature by showing how interface design interacts with user heterogeneity and personalized \textit{recommendation}.


\textit{Post-processing solutions to mis-measurement.} A distinct approach to correcting for measurement error (ratings inflation or consequences of heterogeneous behavior) is to \textit{post-process} the data. Unbalanced explicit training datasets have been documented in the literature and some flattening techniques have been shown to improve recommendation quality~\cite{mansoury2021flatter}. When training algorithms on implicit data, it is natural to sample items that were not consumed as negative feedback \cite{hu2008collaborative}. \citet{stoikovwen} show this approach performs significantly worse than a matrix factorization algorithm trained on explicit like/dislike data. \citet{nosko2015limits} on eBay notice that \textit{missing} ratings are a negative signal of quality, and so dividing a seller's number of positive ratings by the total number of \textit{sales} as opposed to \textit{ratings} provides a stronger signal of seller quality. Outside of a rating context, item response theory \cite{embretson2013item} models are used to simultaneously learn \textit{question difficulty} and \textit{test-taker skill level}, based only on data on how each test-taker performed on each question. Our results point to the importance of using such `rating the rater' methods to process rating data.

%% file: platform.tex
\subsection{The Piki Music App Interface}

\label{sec:platform}

We analyze data from Piki \footnote{The dataset is available at \url{https://github.com/sstoikov/piki-music-dataset} }, a mobile application designed for music discovery. The application presents users with sets of 30-second music video clips in a sequential manner, similar to social media stories. Users are \textit{required} to provide explicit ratings, selecting from options such as ``disliking,'' ``liking,'' or ``super-liking'' for each song. Until a rating is provided, the clip plays in a loop, even across user sessions. Immediately after a rating is provided, the user is presented with the next song. Notably, users do not have access to a search bar and thus have no direct control over the songs they hear. Super-liked songs are saved to a playlist that can be exported to other applications.

The songs presented to users are chosen using one of several recommendation algorithms, which fall into two categories: "random" algorithms that do not use individual user data and "personalized" algorithms that do. Examples of the algorithms used in the application include a randomization technique over a small set of songs to create a dense user-item matrix, a "top-K" system that recommends the most popular songs overall, and a matrix factorization-based recommender. 

As the app focuses on discovery, it has a higher ratio of songs to users than other rating datasets. Therefore, most songs have a "cold start" problem, with a small number of unique ratings. 
In such a setting, generous users who ``like'' all songs indiscriminately may create an artificial bias toward the songs that they have been exposed to. A further concern is that some users may provide feedback randomly rather than according to their true preferences, just to finish a set of songs and get rewards (note that Piki users may choose to receive micro-payments for their ratings).\footnote{Such behavior is a (perhaps extreme) example of the type of heterogeneous and strategic use behavior on many platforms, that might influence both implicit and explicit ratings that they provide.} 
This platform provides several distinct advantages for studying our research questions: (a) it collects explicit ratings, and users must rate songs before continuing; there is thus little selection bias in ratings, conditional on a user being shown the song; (b) it deploys multiple recommendation algorithms of varying levels of personalization, and data is logged in terms of which algorithm led to a song recommendation. These characteristics allow us to focus our attention on the role of user rating heterogeneity, personalized recommendation, and the interface collecting the ratings.



\begin{figure*}[t]
\centering{
		\begin{tabular}{cc}
			\subfloat{\includegraphics[width = 0.1\linewidth]{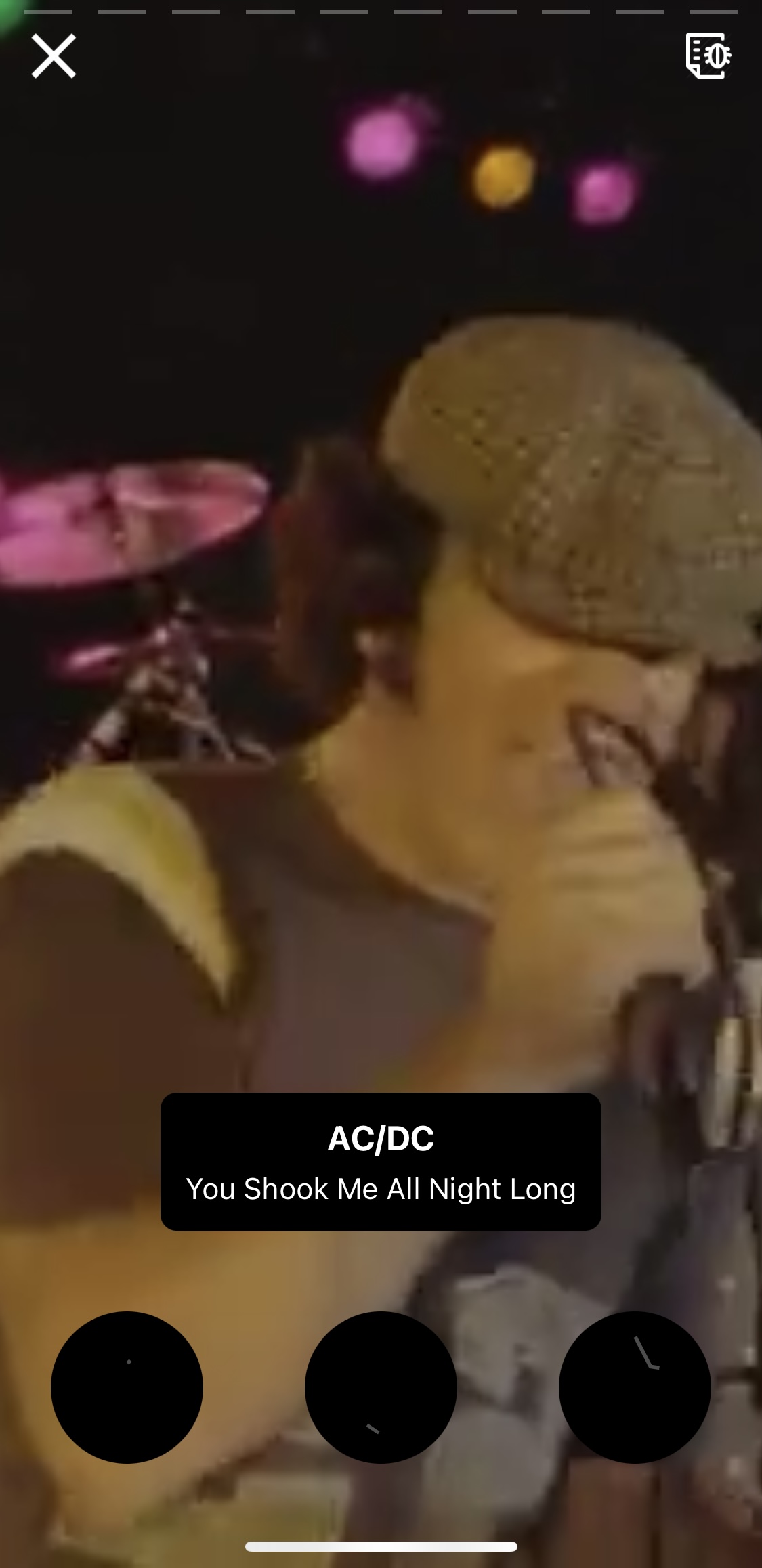}} 
			\subfloat{\includegraphics[width = 0.1\linewidth]{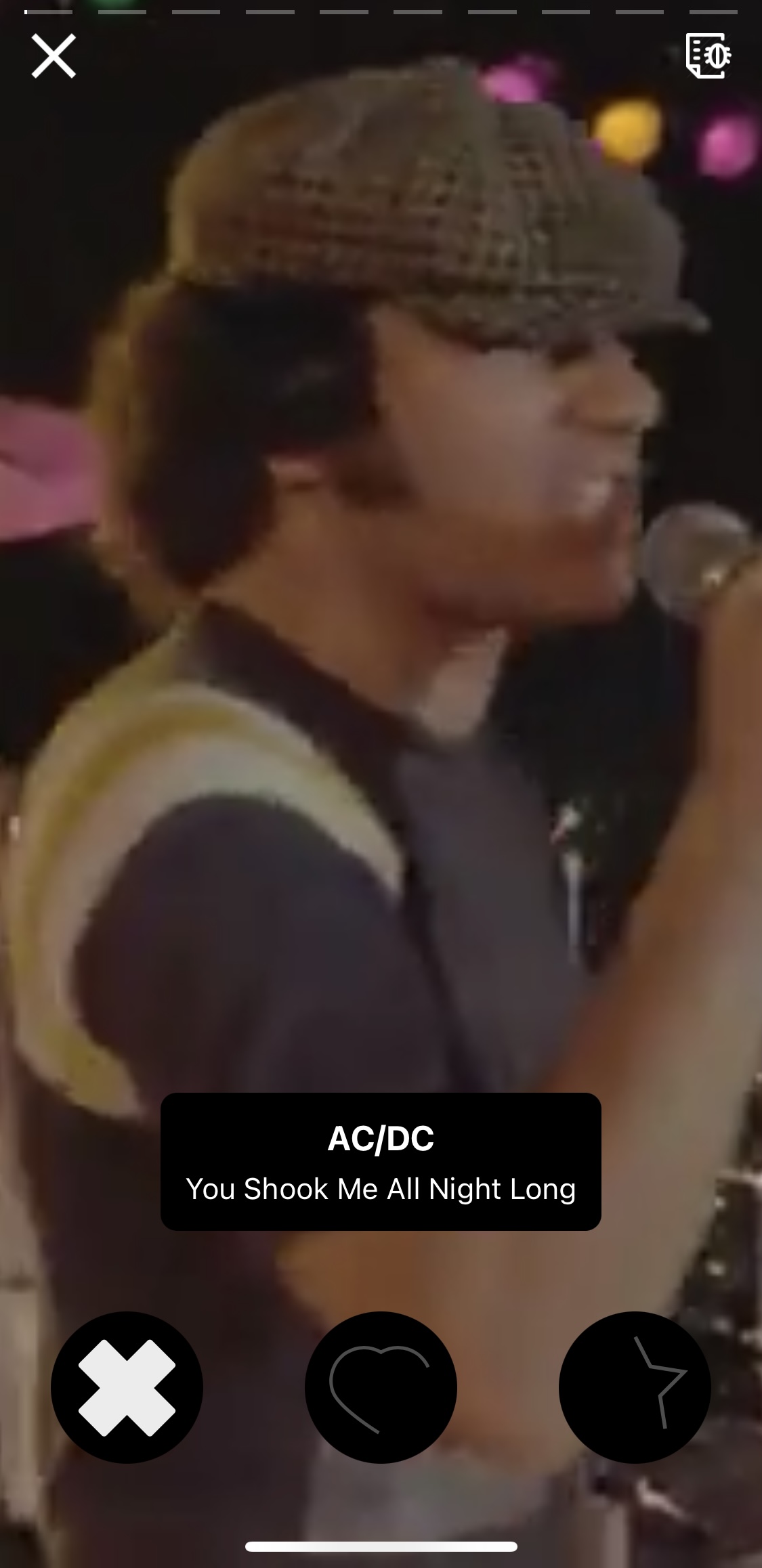}} 
			\subfloat{\includegraphics[width = 0.1\linewidth]{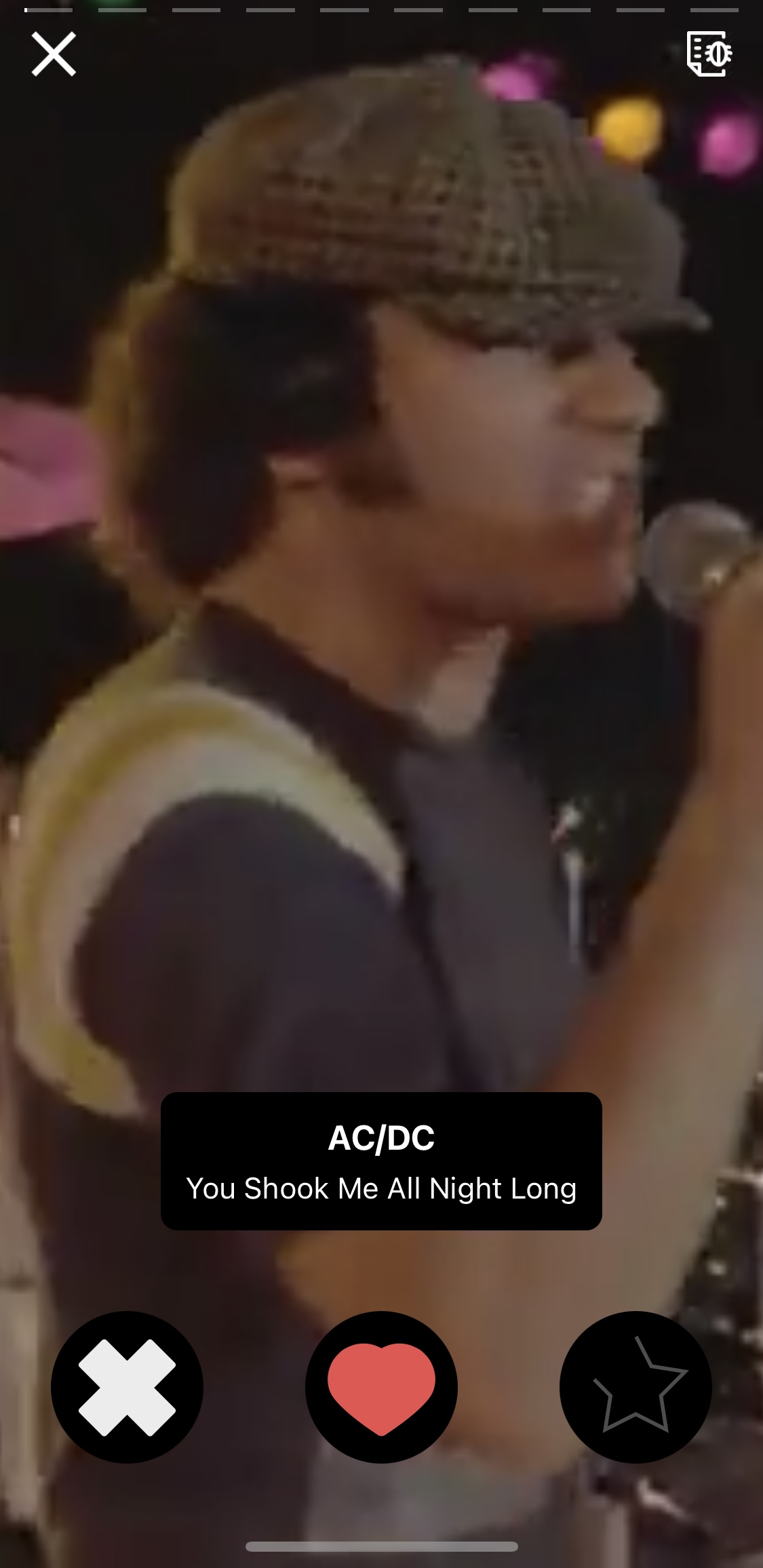}} 
			\subfloat{\includegraphics[width = 0.1\linewidth]{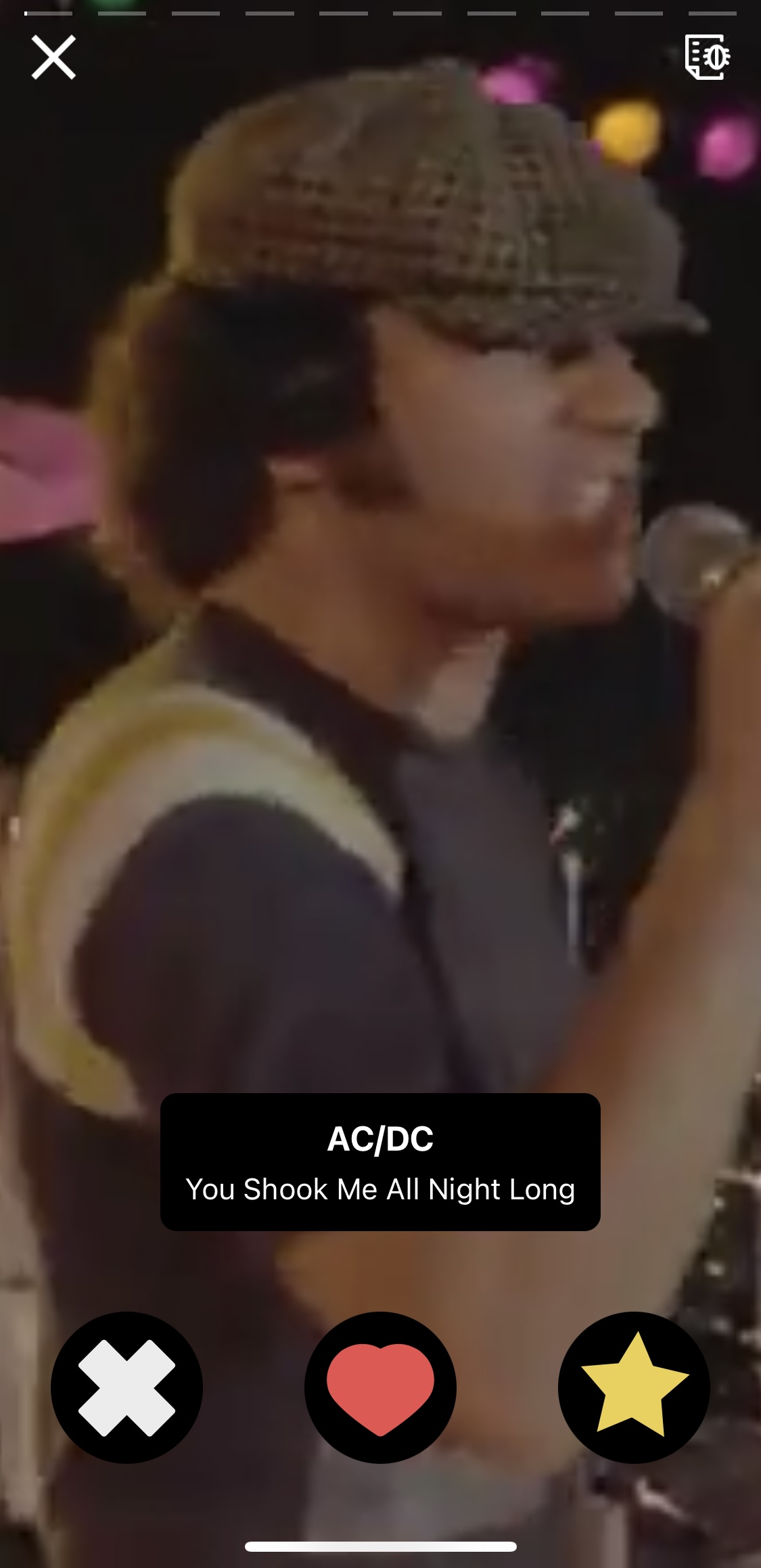}} 
		\end{tabular}
  }
	\caption{The Piki Music App interface. The rating buttons are unlocked after a time period}
	\label{fig:interface}
\end{figure*}


\vspace{-0.15cm}
\paragraph{Piki Interface.} In an effort to mitigate rating inflation and incentivize ratings in line with a user's true opinions, Piki implemented a set of timers on each of the rating options on February 21, 2021. 
Since then, users have not been able to \textit{immediately} rate a song after it begins playing. 
Instead, each of the ``dislike,'' ``like'' and ``super-like'' buttons appear sequentially in that order, several seconds after the song begins playing or the previous button appears. Figure~\ref{fig:interface} displays the application interface.
The timer to unlock the dislike option ensures that users give a chance to each song; the timer to unlock the like option ensures that users are willing to invest extra time in that song; the timer to unlock the superlike option ensures that only raters who invest the significant time will get to save the song. The exact timing is a platform design choice that influences rating behavior and -- as we show -- downstream outcomes. This work first leverages the introduction of the delay timers and then an experiment that further varied the delay time across treatment groups.

%% file: user_heterogeinety_interface.tex
\section{User rating heterogeneity and interface design}
We begin our analysis by analyzing a `natural experiment' induced by a product deployment, as well as an RCT motivated by the natural experiment results. Here, we present results from the two together, as they largely agree with each other. The results show that (a) user rating heterogeneity (how generous they are on average) substantially influences estimated song quality, and (b) interface design can mitigate such heterogeneity.

\subsection*{Natural experiment and RCT design}

\begin{table}[tb]
	\begin{center}
              \begin{tabular*}{\textwidth}{l c r}
             \hspace{3.4cm} \textbf{Users} & \hspace{1.4cm} \textbf{Ratings} &  \hspace{1.65cm}\textbf{Songs} \\
            \end{tabular*}
            \begin{tabular}{p{1.5cm}|p{1cm} p{1cm} | p{1cm} p{1cm} | p{2cm} p{2cm}}
			\textbf{Treatment} & All & Filtered & All & Filtered & All & Filtered \\
			\midrule
			a & $429$ & $389$ & $122225$ & $59100$  & $49598$ & $5454$\\
			b & $331$ & $308$ & $77758$ & $38946$  & $35213$ & $5397$\\
			c & $408$ & $375$ & $92999$ & $43633$  & $42615$ & $5418$\\
            \bottomrule
		\end{tabular}
	\end{center}
	\caption{A total number of users, ratings, and songs (all and filtered) for each treatment group in the RCT.}
	\label{tab:stats}
\end{table}

 \paragraph{Natural experiment: introduction of timers.} Prior to February 21, 2021, users could rate a song as soon as the video started playing; we refer to this data as \emph{Pre-timers dataset}. After the launch (and for at least the 6 months following the change): the dislike button became active after 3 seconds, the like button after 6 seconds, and the superlike button after 12 seconds; the data collected during this period is referred to as the \emph{Post-timers dataset}. We analyze data for the six months before and after the introduction of the timers, filtering out users and songs with less than 10 ratings each. The Pre-timers dataset includes \num[round-precision=0]{396} users, \num[round-precision=0]{8592} songs, and \num[round-precision=0]{141096} overall ratings, while the \emph{post-timers dataset} includes \num[round-precision=0]{1071} users, \num[round-precision=0]{11536} songs, and \num[round-precision=0]{100818} overall ratings. 

\paragraph{RCT design.} The test was conducted with three treatment groups, each of which had a different time delay before the "like" rating could be unlocked. Similar to the post-timer period in the natural experiment, dislike lock time is always 3 seconds, and the superlike lock time is always 12 seconds. In Treatment a, the like button is unlocked after 3 seconds, i.e., at the same time as the dislike button. In Treatment b, after 6 seconds (same as in the post-timer period and before the experiment). In Treatment c, after 9 seconds. We filtered songs and users who provided less than 10 ratings during the experimental period.  
User profiles were randomly allocated to one of the three treatment groups, with Treatment A and Treatment C each receiving $\frac{3}{8}$ of the user profiles and Treatment B receiving $\frac{2}{8}$. Table~\ref{tab:stats} shows the number of users, songs, and ratings for each treatment group before and after data filtering..

\subsection*{Results}

  \begin{figure}[bt]
  \vspace{-0.2cm}
 	\centering
\subfloat[][(Natural Exp) Fraction of ratings]{
 		\includegraphics[width=.32\textwidth]{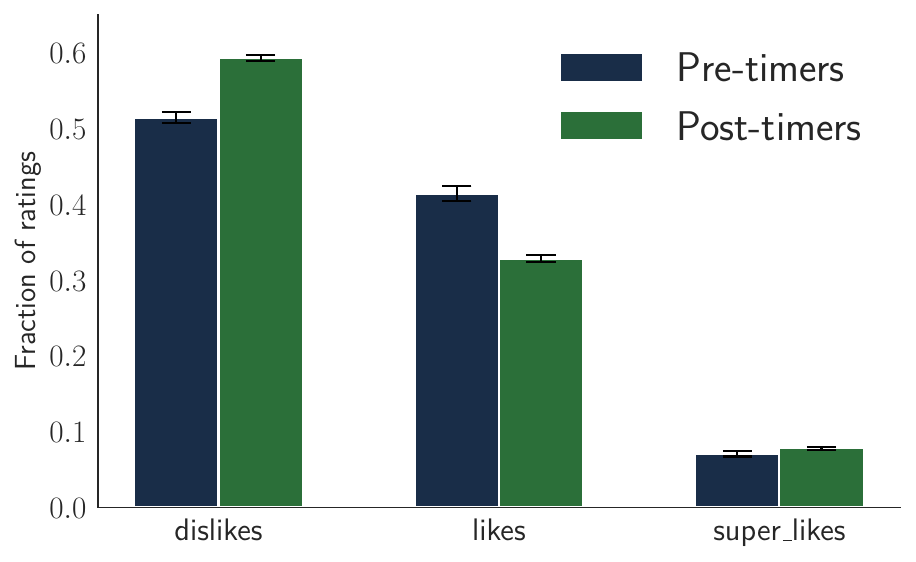}
    \label{fig:mean_rating_before_after}
}
 	\hfill
\subfloat[][(Natural Exp) Mean user scores]{
 		\includegraphics[width=.32\textwidth]{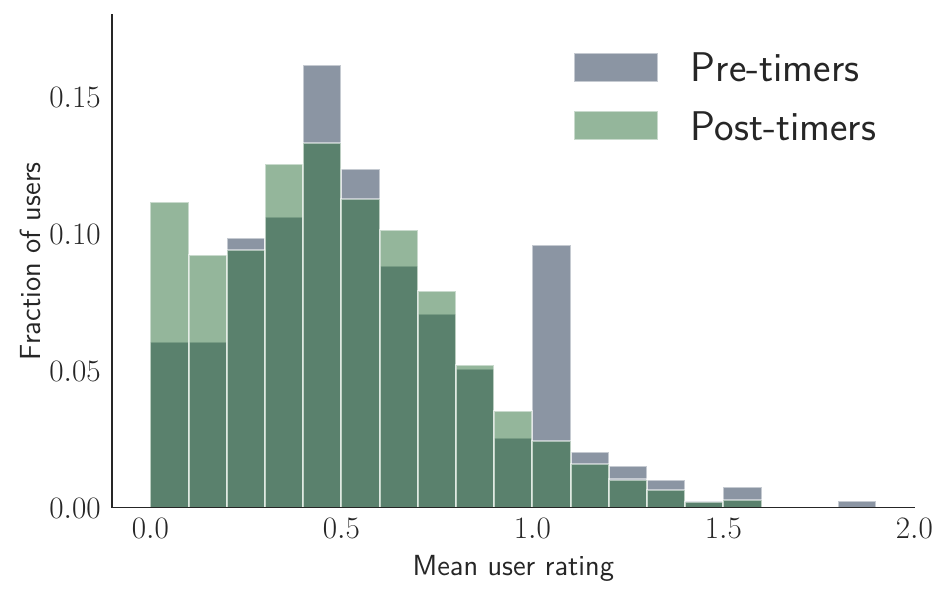}
    \label{fig:user_score_hist}
} 	\hfill
\subfloat[][(RCT) Fraction of ratings ]{
 		\includegraphics[width=.32\textwidth]{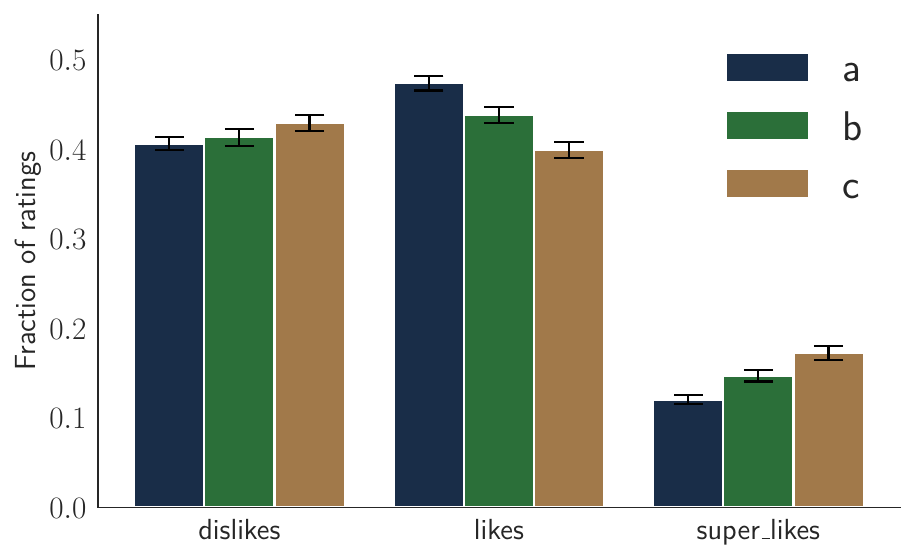}
    \label{fig:mean_rating_exp_all}
}
 	\caption{User rating behavior with various delay timers settings. (a) In the natural experiment, fraction of responses (aggregated first within each user, then across users) for each option (dislikes, likes and superlikes). (b) In the natural experiment, histogram of mean user scores (dislike = 0, like = 1, super-like = 2). (c) Same as (a), except for each RCT treatment. We find that the introduction of timers substantially and statistically significantly changed user rating behavior. The primary effect is making users more negative, since giving a `like' or `super-like' requires waiting additional time. Additionally, user behavior is `more smooth' -- `generous' users who before the change clicked like for every song no longer do. In the RCT, a larger delay on the "like" button decreases the fraction of likes. Unlike the natural experiment, however, the substitution from likes is to \textit{super-likes} with longer timers. In other words, after already waiting 9 seconds for the like button to show up in Treatment c, users are more willing to wait the extra three seconds to instead give a "super-like." }
 	\label{fig:ratings}
 \end{figure}

\paragraph{There is substantial heterogeneity in user rating behavior.} Users behave qualitatively different from each other both before and after the introduction of timers, and the effect cannot be explained by sampling variance alone. For example, \Cref{fig:user_score_hist} shows that raters differ in the mean ratings they provide items. In particular, especially before the change, there are users who would ``like'' every song they are shown, while other users are far more picky.

 {\renewcommand\normalsize{\tiny}%
\normalsize
\begin{table}[tbh]
\input{plots_nikhil/naturalexperiment_likeregression.tex}
\caption{For the natural experiment, regression results for a \textit{single} user-song rating (i.e., either a 0, 1, or 2) versus the average other ratings by that user or for that song. Perhaps surprisingly, this rating is more strongly associated by the \textit{user's} average behavior than the \textit{song's} average rating, especially before the interface design change. All dependent variables are standardized, and so coefficients should be interpreted as the change in the user-song-rating associated with a one standard deviation change in the independent variable. To minimize the effect of any single user or item in this regression and keep the number of ratings by each song and user comparable, we sample at most 100 ratings by any user or for any item. On average, in the post-timer period, each song has 31 ratings; users have on average 29. For the pre-timer period, these numbers are 21 and 22, respectively. Qualitatively similar results emerge for other filtering techniques. }
\label{tab:natexpregression_explainratings}
\end{table}
}

 {\renewcommand\normalsize{\tiny}%
\normalsize
\begin{table}[tbh]
\input{plots_nikhil/naturalexperiment_meanregression.tex}
\caption{For the natural experiment, regression results for a song's mean rating score on a \textit{test} set versus \textit{train set} mean rating scores given by users (who gave the song test-set ratings) or for that song. As in \Cref{tab:natexpregression_explainratings}, mean scores is more strongly associated by the \textit{user} average behavior than the \textit{song's} average rating, especially for songs with low numbers of ratings. All independent and dependent variables are standardized. On average, in the post-timer period, each song has about 16 ratings on average in each of the test and train set; users have on average 22 ratings in the train set. For the pre-timer period, these numbers are 11 and 17, respectively. (Note that the song numbers are exactly half that of \Cref{tab:natexpregression_explainratings}, as the test/train split is stratified by song). Qualitatively similar results emerge for other filtering techniques.}
\label{tab:natexpregression_multiple_ratings}

\end{table}
}

\paragraph{A user-song rating is more a function of which users rate a song than the song itself.} As discussed in the introduction, such ratings behavior heterogeneity may affect items, whose quality estimates depend on the ratings users provide.

We first analyze a regression as shown in Table~\ref{tab:natexpregression_explainratings} to predict a \textit{single} user-item rating as a function of that user's mean rating to other items, and that item's mean ratings by other users. We find that the user's average behavior has a stronger correlation with the individual rating than does the item's average quality as rated by other users. Specifically, in the post-timer period of the natural experiment, a one standard deviation increase in the mean user rating is associated with a \num{0.279} increase in the individual rating, while a one standard deviation increase in the mean item rating is only associated with a \num{.114} increase. Qualitatively similar results emerge for other exact regression specifications.\footnote{For example, one concern is that is the median user rates more songs than the median song receives ratings, and so our estimate of the mean user rating is less noisy, driving the result. However, the result does not change with different filtering/limitations of the number of ratings used for each user and song to calculate the corresponding dependent variables. Furthermore, the interaction terms between the mean ratings and the counts indicate that such an effect is second order. } The RCT yields qualitatively similar results.

One limitation of analyzing a \textit{single} rating is that -- while a single rating may be more a function of the user than the item -- effects may cancel out after multiple ratings, and so an item's estimated quality may indeed be reflective of their true quality. 
To study this effect further,
we employed a procedure where we randomly divided the dataset into two groups, stratifying by both before/after the introduction of timers and song to ensure that an equal number of each song's ratings in each period were present in each group. For each song, we further calculate \textit{mean user rating train} -- the mean of the users' mean train set scores, for users who rated that song in the test set. Finally, we regress \textit{mean user rating train} and the song's mean rating in the training set to the song's mean rating in the ``test'' dataset, alongside controls. Conceptually, this specification tests consistency: is a song's mean score after multiple ratings (in the test dataset) more consistent with the behavior of users who gave those ratings or that song's score as rated by other users?\footnote{Any empirical estimate of true item quality is a function of the data generating process (e.g., user interface) and estimation method, and so cannot be used as (even approximate) ground truth of quality in papers that seek to change the interface. \citet{garg2021designing} address this challenge by using a revealed preference `objective' notion of quality independent of the user interface: whether the freelancer was hired again by the same client. Such a measure is not available in our setting. We thus use internal consistency within a treatment group on a train-test set, as the measure that approximates `consistency to ground truth.'}

Our analysis indicates that a mean song's score remains significantly correlated with user behavior, rather than the quality of the song as assessed by other users, even after tens of ratings, as shown in Table~\ref{tab:natexpregression_multiple_ratings}. For instance, in the post-timer period, these correlations are \num{.478} and \num{.3}, respectively (with all independent variables standardized). Furthermore, the effect lessens but does not disappear for songs with many train set ratings: analyzing the interaction terms based on the number of song and user ratings in the train set, respectively, a one standard deviation increase in the number of train set song ratings (about  14 ratings) approximately equalizes the respective coefficients (however, an increase in the number of train set user ratings also increases the association with the mean train set users' ratings\footnote{The standard deviation in number of train set user ratings is about 5, and so the respective coefficient on a ``per-rating'' basis is $\frac{.0432*14/5}{.198}\approx .6$ that of the interaction term on number of user ratings.}).

Together, these results establish that ratings are not a reliable estimate of item quality: they fail even basic \textit{consistency} checks, being more strongly self-consistent with user behavior than with item quality as measured by ratings.





\paragraph{Ratings interface induces user behavior change.} Finally, our findings show that the ratings interface has a significant impact on user behavior; \Cref{fig:mean_rating_before_after} indicates that the overall number of ``dislikes'' rose on the platform, as expected -- users now have to spend additional time listening in order to give a ``like'' or ``super-like.'' The results from the RCT in \Cref{fig:mean_rating_exp_all} replicate this finding. These overall changes also translate to user-level changes. As shown in \Cref{fig:user_score_hist}, the change was particularly effective in substantially reducing the number of generous users who gave a ``like'' to every selected song -- from about 10\% of users to about 3\%.

These user-level heterogeneity changes further increase how informative ratings are about items as opposed to the users who rated them. Specifically, we observe that the introduction of timers had a noticeable effect on the coefficients in the regression for a single rating. Before the introduction of timers, in the regression for a single rating for a song by a user, the coefficients on the mean user's ratings for other songs is \num{0.278} versus \num{.065} for the mean song's ratings by other users; after timers, these numbers are \num{0.279} and  \num{.114}, respectively, and the interaction between the Pre-Post indicator and mean song rating by others is statistically significant.
These findings highlight the importance of interface design, especially when user behavior is heterogeneous.

\section{Personalized recommendations and ratings}
\label{sec:personalized}

The previous section studied the interaction between user rating behavior and song quality estimates. Next, we analyze how\textit{ personalized recommendations} affect ratings and song quality estimates. At a high level, it is well known that personalized recommendation algorithms can impact item quality estimates. For instance, taking sample means from adaptively collected data, as in a multi-armed bandit, can introduce a negative bias \cite{nie2018adaptively}. Additionally, marketplaces (including the one we study) typically employ \textit{multiple} recommendation algorithms in parallel, such as personalized home screens, search results, "Trending" items, and exploration-motivated randomized recommendations. The coexistence of multiple recommendation algorithms creates a complex ecosystem in terms of how items are recommended to users: if an item is only shown to users as a result of hyper-personalization, its resulting quality estimates will be high; on the other hand, if the same item is only shown to random users regardless of their preferences, its resulting quality estimates will be low.

In other words, the extent of personalization, particularly if this is heterogeneous across items (such as some items are shown in "Trending" but others are not), affects item quality estimates. (Note that such effects can also occur with a single personalization algorithm, based on the quality of recommendations it can provide using each item.)

In this section, we quantify such effects: how do personalized recommendations affect ratings and item quality estimates? We can do so, as a unique feature of the Piki dataset is that we know whether the song was shown as a result of a personalized recommender or at random. 




   \begin{figure}[tbh]
  \vspace{-0.2cm}
 	\centering
  \subfloat[][Histogram of song score]{
 		\includegraphics[width=.30\textwidth]{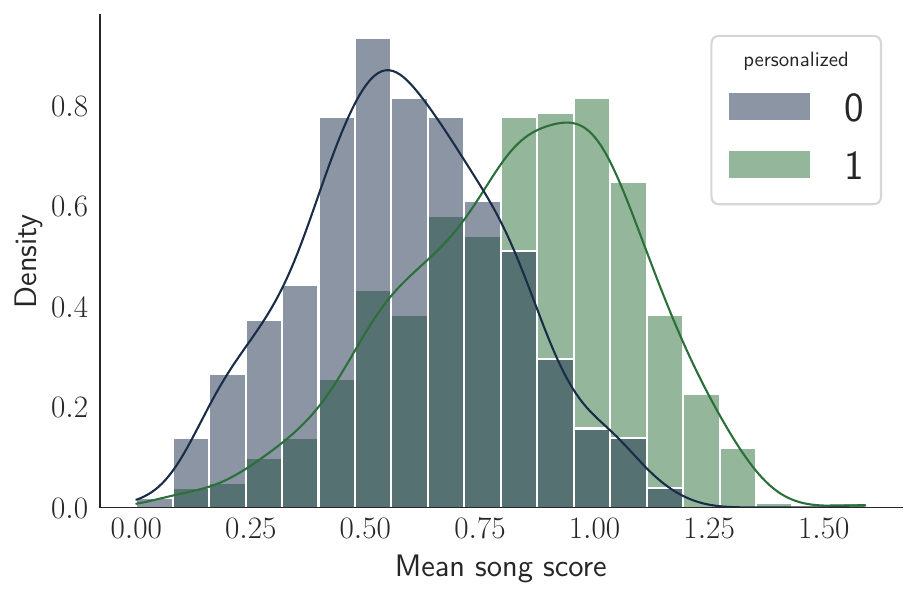}
    \label{fig:exp_personalizedrandomhist}
}
 	\hfill
\subfloat[][Random vs personalized score]{
 		\includegraphics[width=.32\textwidth]{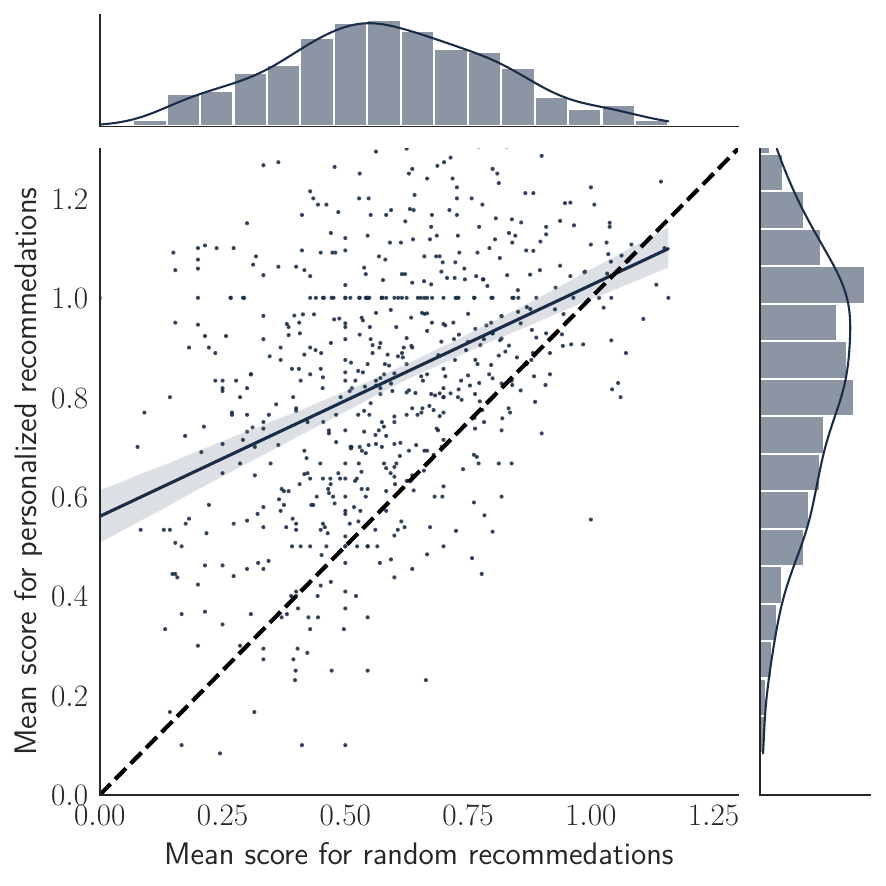}
    \label{fig:exp_personalizedvsrandom}
}
\hfill
\subfloat[][Frac. personalized vs mean score]{
 		\includegraphics[width=.32\textwidth]{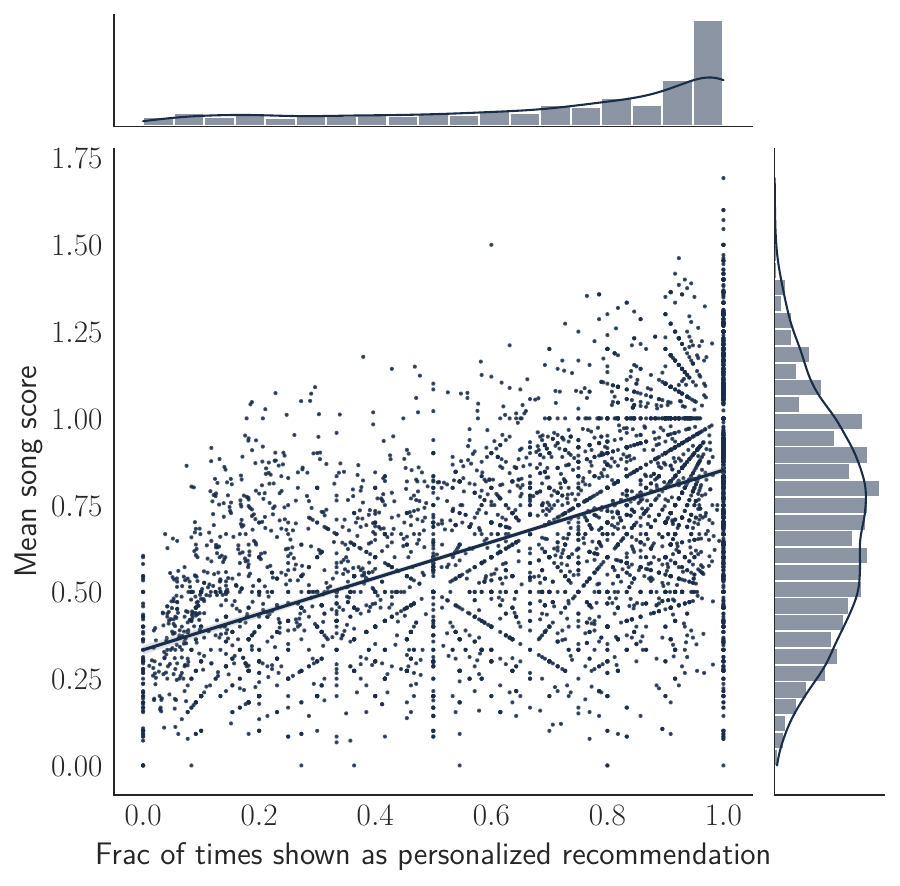}
    \label{fig:exp_personalizedvslike}
}
 	\caption{The relationship between ratings and whether a song was shown due to \textit{personalized} recommendation. (a) Histogram of mean song scores after each type of recommendation. (b) At an individual song level, the relationship between its mean score under personalized and random recommendation, respectively. (c) For each song, the relationship between the fraction of times it is shown under personalized recommendation (out of all the times it is shown) versus its overall mean song score.}
 	\label{fig:ratings}
 \end{figure}

We start with a basic observation: ratings after a song is shown due to a personalized recommendation are higher than after a song is shown due to randomized recommendation.\footnote{Note that users are not aware of why a particular song is shown, and so the effect is not due to a placebo effect of the user `expecting' to like such a personalized recommendation more.} For songs with more than 10 ratings due to each of random and personalized recommendation, \Cref{fig:exp_personalizedrandomhist} shows the histogram of mean song scores under each of personalized and random recommendation. \Cref{fig:exp_personalizedvsrandom} then shows the individual-song level relationship (for each song, the mean score under personalized and random recommendation, respectively). As expected, scores are higher under personalized recommendation, indicating the algorithms are working. There is a generally positive relationship at the song level, but perhaps weaker than expected. 

However, this observation -- that personalization leads to higher ratings -- induces the next observation: estimated song quality is \textit{also} a function of whether a song is shown more often as a result of personalized or random recommendation. (Note that heterogeneity in how often a song is shown via personalized recommendation can be a function of randomness, system design choices, and historic ratings data).  As an initial view of this effect, consider \Cref{fig:exp_personalizedvslike}: the higher the fraction of times that an item is shown due to personalized recommendation (versus randomized), the higher its estimated song score. Of course, causality can be in either direction: a higher mean song score may mean that a personalized recommender is more likely to recommend that song, and being shown through personalized recommendation results in higher mean scores. 

To further analyze the effect, we run a regression as follows: mean song scores on a test set, versus user mean ratings and that song's score on a train set, further adding as a covariate a term for the fraction of times that a song was recommended via personalization in the test set. This fraction has a strong positive association with the average song rating in the test set, even controlling for song quality as determined in the training set. In other words, the more that a song was shown via personalized recommendation in the test set, the higher its test set song quality estimation score, even controlling for train set score. This analysis shows that the frequency with which songs are displayed through personalized recommendations, even after conditioning on true quality, can significantly impact the platform's learning about items.

%% file: plots_nikhil/naturalexperiment_likeregression.tex
\begin{center}
\begin{tabular}{lclc}
\toprule
\textbf{Dep. Variable:}                                      & user\_song\_rating & \textbf{  R-squared:         } &     0.225   \\
\textbf{Model:}                                              &        OLS         & \textbf{  Adj. R-squared:    } &     0.224   \\
\textbf{Method:}                                             &   Least Squares    & \textbf{  F-statistic:       } &     917.6   \\
\textbf{No. Observations:}                                   &        31692       & \\
\bottomrule
\end{tabular}
\begin{tabular}{lcccccc}
                                                             & \textbf{coef} & \textbf{std err} & \textbf{t} & \textbf{P$> |$t$|$} & \textbf{[0.025} & \textbf{0.975]}  \\
\midrule
\textbf{Pre\_Post[post\_timers]}                             &       0.4231  &        0.008     &    52.415  &         0.000        &        0.407    &        0.439     \\
\textbf{Pre\_Post[pre\_timers]}                              &       0.4272  &        0.010     &    41.280  &         0.000        &        0.407    &        0.447     \\
\textbf{personalized}                                        &       0.0918  &        0.010     &     9.538  &         0.000        &        0.073    &        0.111     \\
\textbf{mean\_user\_rating\_others}                          &       0.2788  &        0.004     &    76.931  &         0.000        &        0.272    &        0.286     \\
\textbf{mean\_user\_rating\_others:Pre\_Post[T.pre\_timers]} &      -0.0100  &        0.008     &    -1.259  &         0.208        &       -0.026    &        0.006     \\
\textbf{mean\_song\_rating\_others}                          &       0.1144  &        0.004     &    28.431  &         0.000        &        0.107    &        0.122     \\
\textbf{mean\_song\_rating\_others:Pre\_Post[T.pre\_timers]} &      -0.0499  &        0.008     &    -6.320  &         0.000        &       -0.065    &       -0.034     \\
\textbf{user\_ratings\_count}                                &      -0.0042  &        0.003     &    -1.283  &         0.200        &       -0.011    &        0.002     \\
\textbf{mean\_user\_rating\_others:user\_ratings\_count}     &       0.0199  &        0.003     &     6.009  &         0.000        &        0.013    &        0.026     \\
\textbf{song\_ratings\_count}                                &       0.0136  &        0.004     &     3.488  &         0.000        &        0.006    &        0.021     \\
\textbf{mean\_song\_rating\_others:song\_ratings\_count}     &       0.0348  &        0.003     &    10.128  &         0.000        &        0.028    &        0.042     \\
\bottomrule
\end{tabular}
\end{center}


%% file: plots_nikhil/naturalexperiment_meanregression.tex
\begin{center}
\begin{tabular}{lclc}
\toprule
\textbf{Dep. Variable:}                                        & mean\_song\_rating\_test & \textbf{  R-squared:         } &     0.327   \\
\textbf{Model:}                                                &           OLS            & \textbf{  Adj. R-squared:    } &     0.322   \\
\textbf{Method:}                                               &      Least Squares       & \textbf{  F-statistic:       } &     59.53   \\
\textbf{No. Observations:}                                     &            1112          & \\
\bottomrule
\end{tabular}
\begin{tabular}{lcccccc}
                                                               & \textbf{coef} & \textbf{std err} & \textbf{t} & \textbf{P$> |$t$|$} & \textbf{[0.025} & \textbf{0.975]}  \\
\midrule
\textbf{Pre\_Post[post\_timers]}                               &      -0.0082  &        0.029     &    -0.279  &         0.781        &       -0.066    &        0.049     \\
\textbf{Pre\_Post[pre\_timers]}                                &      -0.0013  &        0.048     &    -0.027  &         0.978        &       -0.095    &        0.092     \\
\textbf{mean\_user\_rating\_train}                             &       0.4775  &        0.030     &    15.860  &         0.000        &        0.418    &        0.537     \\
\textbf{mean\_user\_rating\_train:Pre\_Post[T.pre\_timers]}    &      -0.0296  &        0.057     &    -0.518  &         0.604        &       -0.142    &        0.082     \\
\textbf{mean\_song\_rating\_train}                             &       0.2999  &        0.032     &     9.505  &         0.000        &        0.238    &        0.362     \\
\textbf{mean\_song\_rating\_train:Pre\_Post[T.pre\_timers]}    &      -0.0774  &        0.056     &    -1.377  &         0.169        &       -0.188    &        0.033     \\
\textbf{count\_song\_ratings\_train}                           &       0.0200  &        0.027     &     0.744  &         0.457        &       -0.033    &        0.073     \\
\textbf{mean\_song\_rating\_train:count\_song\_ratings\_train} &       0.1980  &        0.034     &     5.905  &         0.000        &        0.132    &        0.264     \\
\textbf{count\_user\_ratings\_train}                           &       0.0598  &        0.027     &     2.208  &         0.027        &        0.007    &        0.113     \\
\textbf{mean\_user\_rating\_train:count\_user\_ratings\_train} &       0.0432  &        0.024     &     1.775  &         0.076        &       -0.005    &        0.091     \\
\bottomrule
\end{tabular}
\end{center}


%% file: discussion.tex
\section{Conclusion}
\label{sec:discussion}

This paper presents an analysis of the impact of heterogeneous user rating behavior and personalized algorithms on the collection of ratings, using data from Piki, a music discovery platform. 
%
We analyzed the results of a randomized controlled trial (RCT) on the Piki platform.
The users were randomly assigned to different treatment groups, each with a different waiting time before they could "like" a song.
Our research indicates that carefully designing the user interface can influence user behavior in a way that mitigates rating inflation. This leads to more informative ratings and affects downstream personalized recommendations. 
%
%
%
%
Our work has several implications for platform rating system design and future work. First, and most simply, item quality estimates as a function of ratings should not be viewed as sole functions of item quality; as we show, heterogeneous user behavior and personalized recommendation dynamics also play a large role. Second, platforms should consider several interventions to mitigate inflation. Here, we study rating interface changes, but algorithmic post-processing could also play a role. 

\textbf{Acknowledgements:} This work is partially sponsored by Meta Research Award, Jacobs Technion-Cornell Institute at Cornell Tech, and by Zuckerman Foundation.

%% file: appendix.tex
\section{Supplementary information and analyses}

\input{appendix_regressiontables}

\begin{figure}[tbh]
\centering
\includegraphics[scale=0.5]{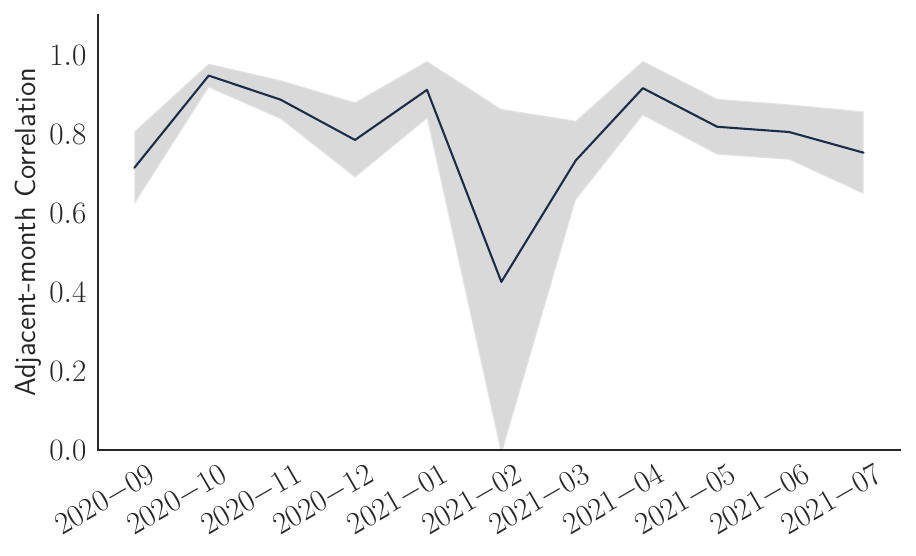}
\caption{Correlation between each user's mean rating in two consecutive months over a period of 16 months. The dip corresponds to the launch of the timers, which heterogeneously changed user behavior -- and so a user's mean rating before and after timers is lower than just considering two before-timer periods or two after-timer periods.}
\label{fig:split_corr}
\end{figure}

     {\renewcommand\normalsize{\small}%
    \normalsize
    \begin{table}[tbh]
    \centering {
    \input{plots_nikhil/experiment_uservariance.tex}}
    \caption{Analogue of \Cref{fig:user_score_hist} for the experiment -- the \textit{variance} in the user score histograms across treatment groups, for \textit{all} recommendations and for \textit{random} and \textit{personalized} recommendations, respectively. The confidence intervals are user-level bootstrapped 95\% CIs of the variance of the user score distribution. }
    \label{tab:exp_score_variance}
    \end{table}
    }

 {\renewcommand\normalsize{\tiny}%
\normalsize
\begin{table}[tbh]
\input{plots_nikhil/experiment_meanregression.tex}
\caption{Exact analogue (without fraction personalized) of \Cref{tab:natexpregression_multiple_ratings} for the RCT. Regression results for a song's mean rating score on a \textit{test} set versus \textit{train set} mean rating scores given by users (who gave the song test-set ratings) or for that song. The mean-song-rating coefficient is approximately the same as the mean-user-rating coefficient, suggesting that both song quality and user behavior contribute to a song's mean rating. We find no significant differences between treatment groups. All independent and dependent variables are standardized. On average, in each treatment, each song has about 12 ratings on average in each of the test and train set; users have on average 23 ratings in the train set. Qualitatively similar results emerge for other filtering techniques.}
\label{tab:experiment_regression_explainratings_multiple}
\end{table}
}

\begin{figure*}[t]
	\center{
		\begin{tabular}{cc}
			\subfloat[Group a]{\includegraphics[width = 0.3\linewidth]
{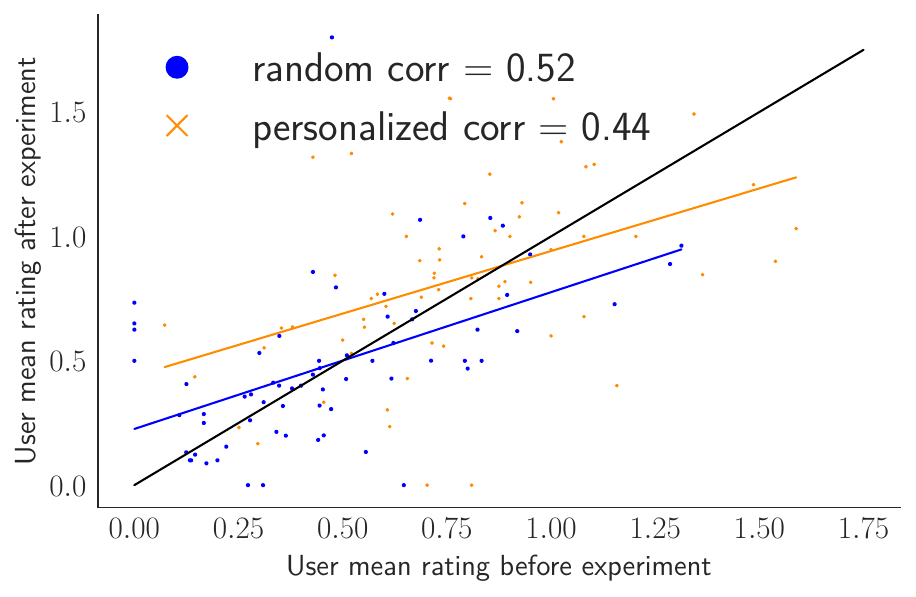}\label{fig:rating_beofre_after_a}} &
			\subfloat[Group b]{\includegraphics[width = 0.3\linewidth]{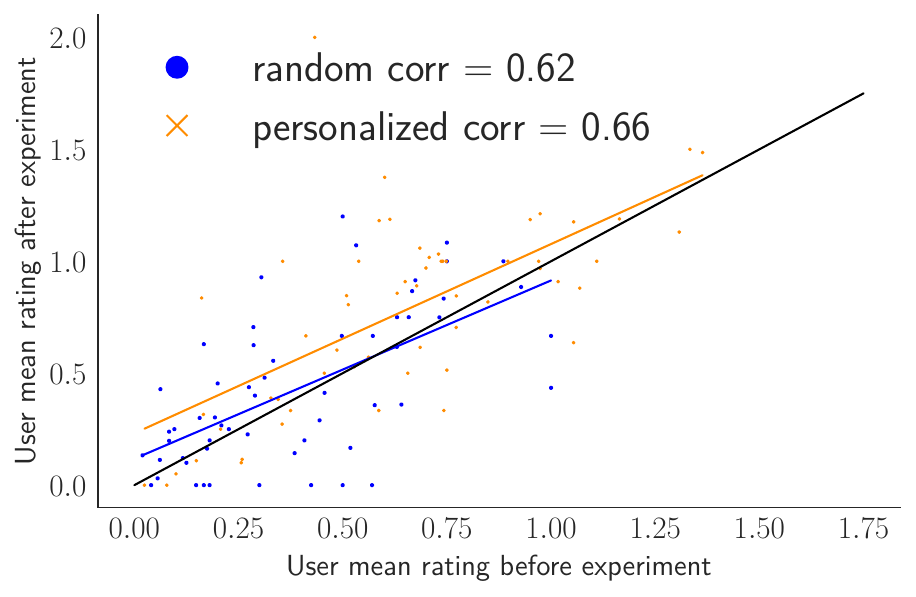}\label{fig:rating_beofre_after_b}}
			\subfloat[Group c]{\includegraphics[width = 0.3\linewidth]{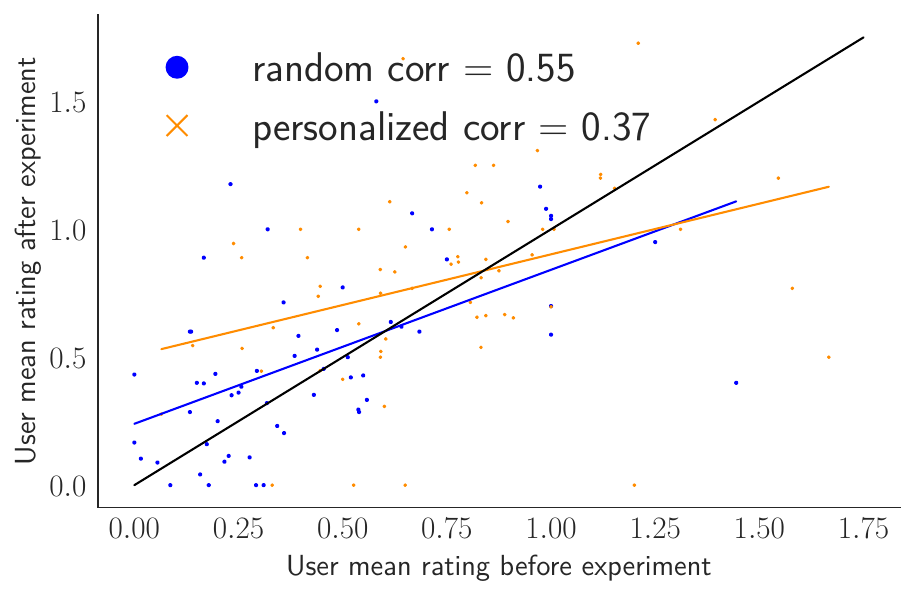}\label{fig:rating_beofre_after_c}}

		\end{tabular}
		}
	\caption{The correlation between mean user rating before and after the experiment in each treatment group. Group b has the same timer delays as the pre-experiment period, and so, unsurprisingly, induces  the most consistent user behavior at the individual level. }
	\label{fig:rating_beofre_after}
\end{figure*}

\begin{figure*}[tbh]
	\center{
		\begin{tabular}{cc}
			\subfloat[Group a]{\includegraphics[width = 0.3\linewidth]
				{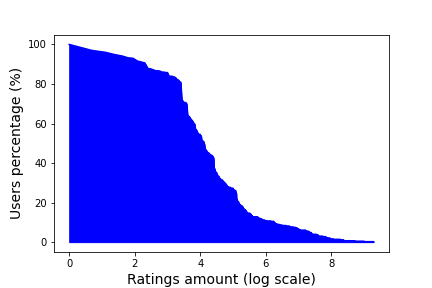}\label{fig:retension_a}} &
			\subfloat[Group b]{\includegraphics[width = 0.3\linewidth]{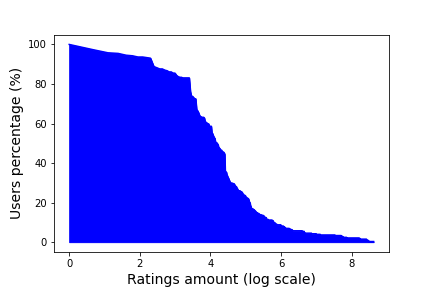}\label{fig:retension_b}}
			\subfloat[Group c]{\includegraphics[width = 0.3\linewidth]{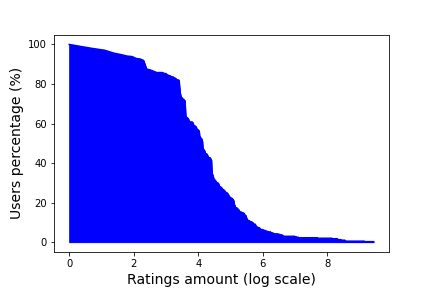}\label{fig:retension_c}}

		\end{tabular}
		}
	\caption{Users' percentage per rating number in log scale for each treatment group. There is no important difference between treatment groups in terms of user retention.}
	\label{fig:retension}
\end{figure*}

\paragraph{\textbf{Personalized songs ratings are affected the most from changing the interface}}
Figure~\ref{fig:rating_beofre_after} shows the correlation between mean user rating before and after the experiment in each treatment group.
We examine two different rating types: ratings for personalized songs and random songs.
As demonstrated, the correlation in Group b is the highest among the other groups for personalized song rating; this is due to the fact that we had the same configuration as Group b before the experiment.
The correlation of the personalized songs rating is lower than the random songs rating in the other treatment groups (a and c), where the interface was changed before and after the experiment.

\paragraph{\textbf{Users retention is not affected from changing the interface}}
We compared median ratings per user in each treatment group to examine how the interface influenced user retention.
The median number of ratings per user in Group a is $61$, Group b is $66$, and Group c is $62$.
Furthermore, Figure~\ref{fig:retension} shows the users' percentage per rating number in log scale.
There are no substantial differences between the three treatment groups.
This means that in each group, the same percentage of users give varying number of ratings.
We can conclude that the interface does not have a strong effect on the number of ratings given by users, which means users retention is not affected as well.

\subsection{Simulation: How do ratings affect the quality of personalized recommendation?}
\label{sec:sim}

We now study how rating behavior affects personalized recommendation. In all the above analyses, we primarily study platform \textit{informational} effects -- how does the user interface and personalized recommendations affect the information that the platform has about items. While better signal about song quality may itself be an outcome of interest (and is to the Piki platform), it is not clear that such informational effects have \textit{downstream} implications for users, e.g., via better personalized recommendations. In particular, \citet{krauth2020offline} show via simulation that offline metrics may poorly correlate with {online} metrics in recommendation systems. 

To study these effects for our research questions, we also employ a simulation approach, modifying the RecLab~\cite{krauth2020offline} simulation framework. We calibrate ratings behavior to the experimental data -- each simulated ratings ``interface'' induces a different distribution of item ratings, which are used as inputs to the recommendation algorithm. (Note that we make the simplifying assumption to not capture user \textit{ratings} heterogeneity, while still simulating user \textit{preference} heterogeneity). A simulation allows us knowledge of "ground truth" user preferences and item characteristics.


\subsubsection{Simulation Setup}
RecLab's evaluation process consists of two main components: environments and recommenders. An environment is a set of users and items  (each with associated covariates), and a recommender interacts with the environment iteratively.

At each step of the simulation, a random subset of users is selected. These users are then provided with a recommendation for an item from the initialized recommender. The ground truth user preference for the recommended is a function of the user and item covariates. The user provides a rating as a function of this ground truth preference and the ``interface'' that is being tested (which, as in the above analyses, shifts the relationship between true user preferences and provided ratings). Based on the ratings, the recommender is updated. This process is repeated for many time steps.

\paragraph{Simulation parameters.} 
In the simulation, we construct factors of size $k$ for both users and items. The user factors are sampled from a uniform distribution $U[0,1]$, while the item factors are sampled from a gamma distribution $\Gamma(2,1)$.

\textit{Ratings.} How are ratings simulated? Using the above factors, we calculate the ground truth preference as the dot product of the user and item factors, resulting in:
$ ground\_truth\_preference = user_i \cdot item_i \sim \Gamma(k,1) $, i.e., true preferences follow a gamma distribution with shape parameter $k$ and scale parameter $1$. 
To account for noise in the ratings, we add normal noise to the ground truth preference in the form of continuous $noisy\_preference = ground\_truth\_preference + N(0,\sigma^2)$, where $\sigma$ is a noise coefficient. Finally, we threshold the noisy preference rating to convert it to one of three ordinal ratings as in our empirical context: dislike, like, or superlike.

In different simulations, we vary the thresholds and thus the overall ratings distributions. For example, in one set, we calibrate the thresholds to replicate the ratings distributions in the RCT. 

\textit{Recommendations.} Three different recommendation algorithms were evaluated in each simulation: a random recommender, libFM, and toppop. The random recommender algorithm predicts ratings uniformly at random, while libFM is a factorization machine algorithm implemented in LibFM \cite{rendle2012factorization}. The toppop algorithm recommends the most popular items to every user without personalization. The popularity of each item was determined by its average rating.

Simulation parameters are detailed in \Cref{tab:sim_param}.





\textit{Simulation Experiments} We conducted two simulations, \emph{sim\_exp, sim\_ctld}, each one involved simulating three different interface designs that varied in the timer duration. The treatment groups were designated as a, b, and c in each simulation. The first simulation (\emph{sim\_exp}), closely resembled the experimental design by setting $threshold_1$ and $threshold_2$ according to the experiment results. While
the second simulation (\emph{sim\_ctld}) involved more extreme manipulation of the timer durations, such that Treatment a had a timer duration that favored likes and Treatment c had a timer duration that favored dislikes. (We note that each ``treatment'' is independent of the others, and so the grouping of treatments into \emph{sim\_exp, sim\_ctld} are only for ease of reporting results).

\begin{table}[th]
	\begin{center}
		\begin{tabular}{p{5cm}|p{6cm}|p{2cm}}
			\toprule
			\textbf{Parameter} & \textbf{Description} & \textbf{Value}\\
			\midrule
			$num\_users$& users number & $100$ \\
			$num\_items$& items number & $5000$ \\
			$n\_iter$& iterations number & $1000$ \\
			$ratio\_init\_ratings$& ratio of ratings employed for training the recommender from the entirety ratings & $0.01$ \\
			$rating\_frequency$& ratio of iteration rating from all rating & $0.1$ \\
            $latent\_dim$& factors number (parameter k)  & $5000$ \\
			$\sigma$& noise multiple factor & $0.5$ \\
			${like,dislike,superlike}\_weight$& ratings encoding & ${0,1,2}$ \\
			$thresholds\_name$& name of thresholds array that used to map the continuous rating to {dislike, like, superlike} of the two simulations & $sim\_exp$ or $sim\_ctld$\\
            \bottomrule
		\end{tabular}
        \begin{tabular}{lclc}
        \textbf{Simulation}  &  \textbf{Treatment}   &  \textbf{$threshold_1$}  &  \textbf{$threshold_2$}  \\
        \textbf{$sim\_exp$} &   a  & $0.4028$ & $0.8845$  \\
        &   b  & $0.4276$ & $0.8508$  \\
        &   c & $0.4240$ & $0.8270$  \\
        \textbf{$sim\_ctld$} &   a  & $0.33$ & $0.66$  \\
        &   b  & $0.25$ & $0.5$  \\
        &   c & $0.5$ & $0.75$  \\
        \bottomrule
        \end{tabular}
	\end{center}
	\caption{Simulation parameters }
	\label{tab:sim_param}
\end{table}



\subsubsection{Results}

  \begin{figure}[bt]
  \vspace{-0.2cm}
 	\centering
\subfloat[][Ratings under Random recommendation]{
 		\includegraphics[width=.48\textwidth]{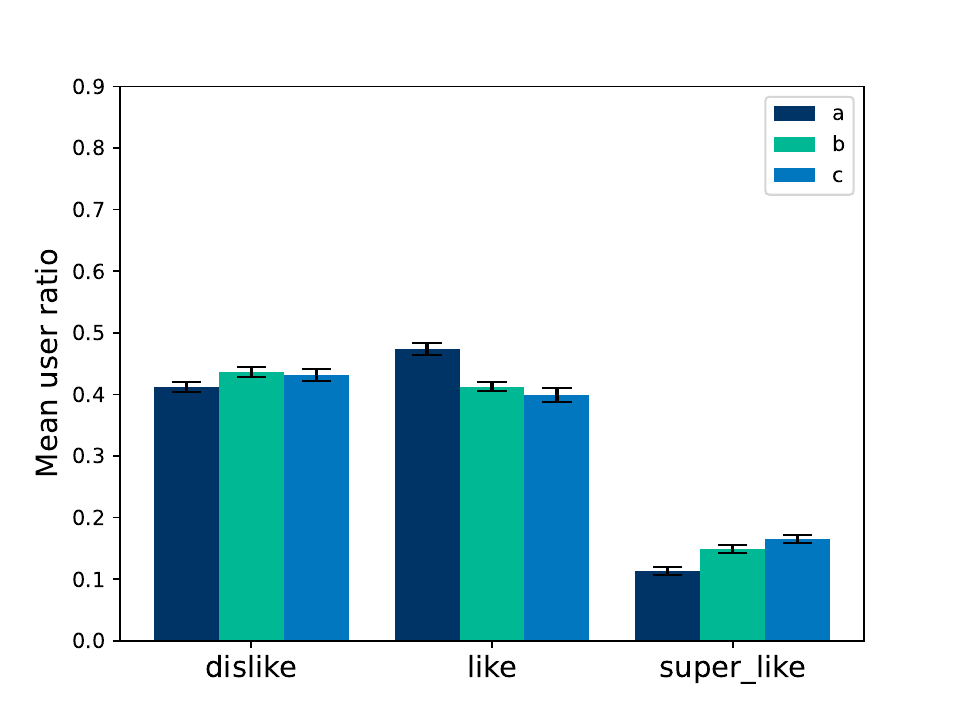}
    \label{fig:mean_rating_sim_random}
}
 	\hfill
\subfloat[][Utility over time, LibFM recommender]{
 		\includegraphics[width=.48\textwidth]{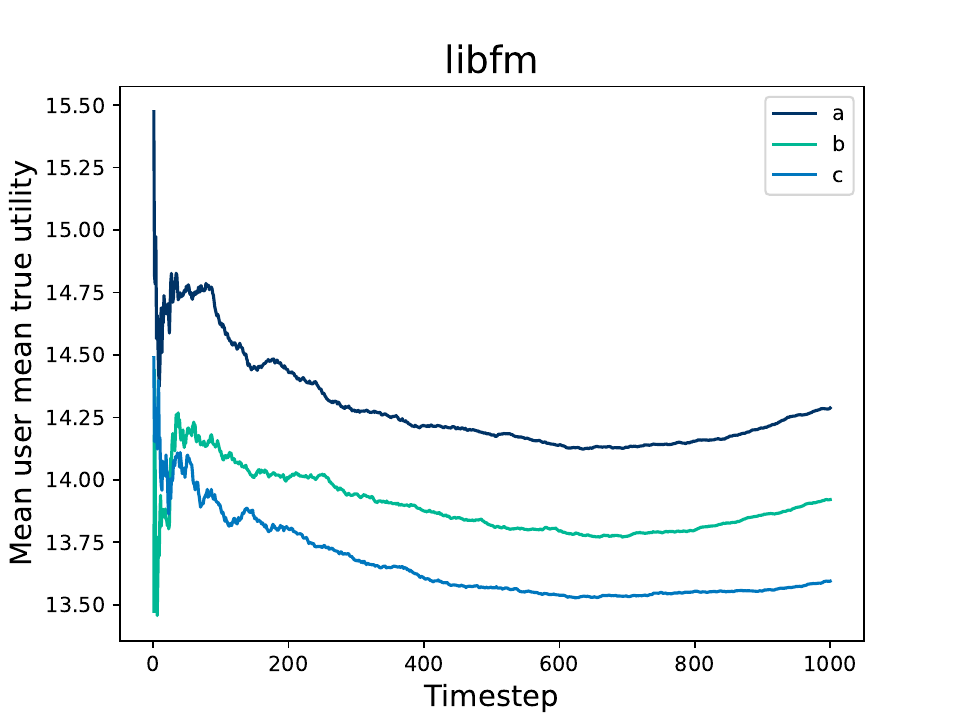}
    \label{fig:mean_rating_sim_exp_libfm}
}
 	\caption{Simulated experiment results. (a) Fraction of responses (aggregated first within each user, then across users) for each option (dislikes, likes and superlikes) using Random recommender. (b) mean user utility over time using LibFM recommender. A simulated experiment follows the experimental conduct as the distribution of both indicates that a has the least dislikes and the least super-likes, whereas c is the most disliked and most super-liked (See Figure~\ref{fig:mean_rating_exp_all}). The results indicate that Treatment a provides the highest level of utility to its users.}
 	\label{fig:sim_exp_res}
 \end{figure}

The results of the simulation calibrated to the experimental data are depicted in Figure~\ref{fig:sim_exp_res}. Specifically, we report: (1) the ratings distribution (dislikes, likes, and superlikes) for under random recommendation, and (2) the mean user utility under libFM recommendations, computed as the average the \textit{ground truth} user preference for items recommended to each user. The results for the remaining recommender algorithms and simulations are provided in the appendix, in Figures \ref{fig:sim_ctld_res}-\ref{fig:sim_utils}. 

\paragraph{Calibration check for the ratings distribution.} Figures~\ref{fig:mean_rating_exp_all} and~\ref{fig:mean_rating_sim_random} illustrate the simulated experiment closely mirrors the experimental conduct (in terms of overall ratings distributions). The distribution of both figures indicate that Treatment a has the least number of dislikes and super-likes, whereas Treatment c receives the most dislikes and super-likes.


\paragraph{Different interfaces lead to different personalized recommendations for users in the long run.} User utility refers to the overall satisfaction of users with the recommendations provided by the recommendation system in this study. In the simulation, the mean user utility is computed by averaging the user's ground truth preference for each item and then averaging the resulting values for items they were recommended. (Note that such ground truth preferences are only available in simulation).

The results indicate that the ratings distribution can have substantial effect on the quality of personalized recommendations. Specifically, in the calibrated simulation, Figure~\ref{fig:mean_rating_sim_exp_libfm} shows that Treatment a has the highest level of user utility over time. Interestingly, this result is in the opposite direction as our offline experimental metrics, where Treatment c seems to provide the highest accuracy quality estimates for items (of course the simulation differs in various aspects, most notably we do not simulate user ratings heterogeneity and so the settings are not directly comparable). Figure~\ref{fig:mean_rating_sim_ctld_libfm} shows similar levels of differences in user utility across simulated treatments.

Together, the results establish the large potential effect that the ratings data has on the quality of personalized recommendation, and that the relationship with quality estimation is complex. A rich avenue for future work is to theoretically characterize how the ratings distribution affects such personalized recommendation, as \citet{garg2019designing,garg2021designing} do for item quality estimation in a non-personalized context.  



  \begin{figure}[bt]
  \vspace{-0.2cm}
 	\centering
\subfloat[][Ratings under Random recommendation]{
 		\includegraphics[width=.48\textwidth]{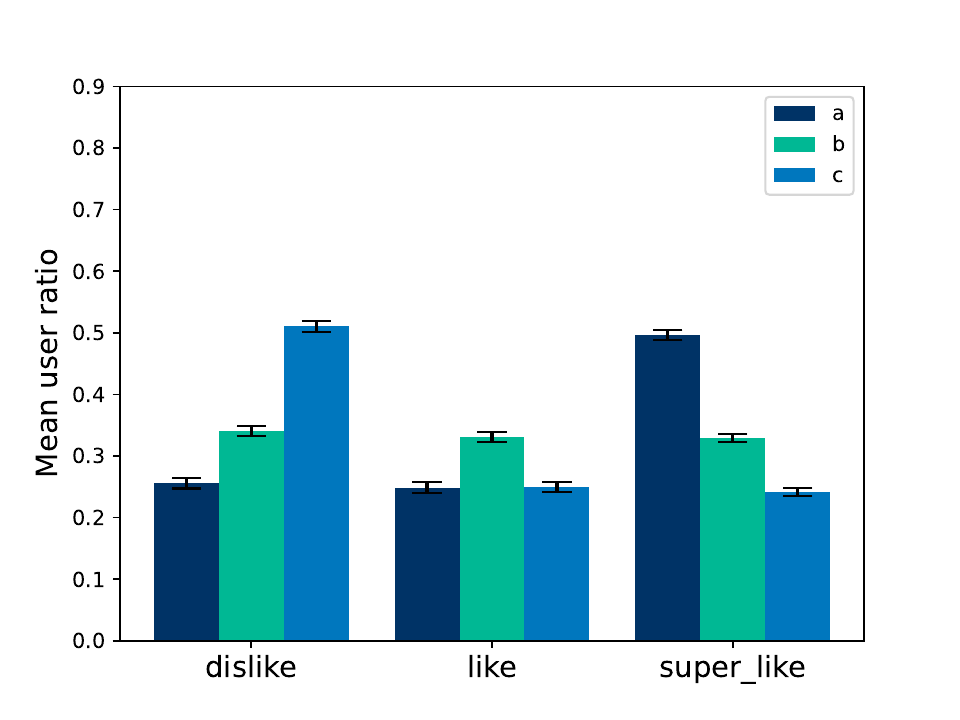}
    \label{fig:mean_rating_sim_ctld_random}
}
 	\hfill
\subfloat[][Utility over time, LibFM recommender]{
 		\includegraphics[width=.48\textwidth]{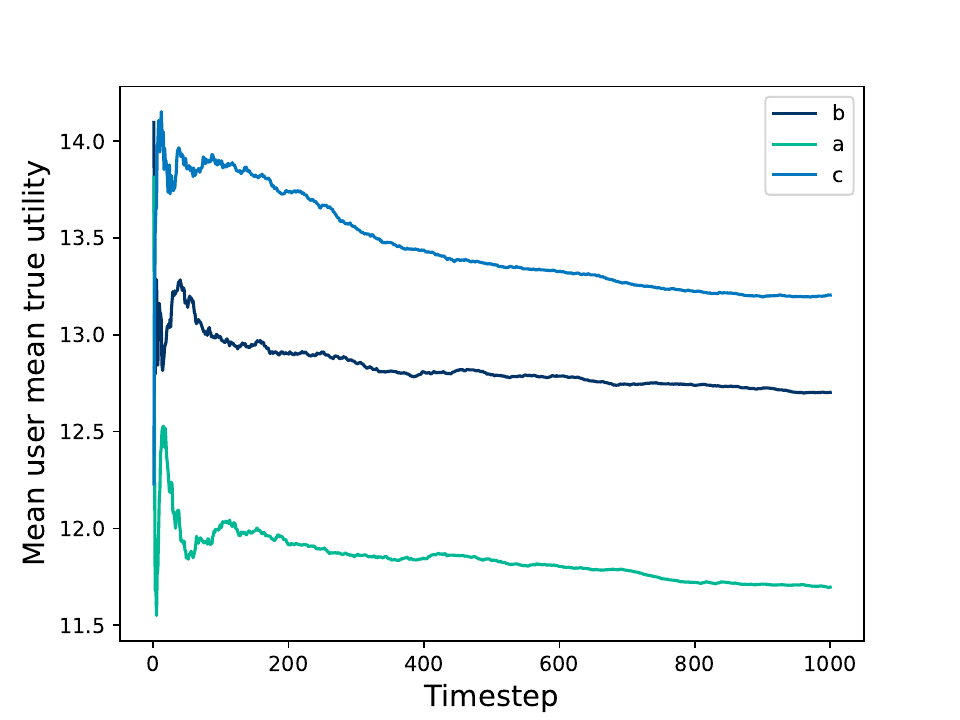}
    \label{fig:mean_rating_sim_ctld_libfm}
}
 	\caption{Controlled simulation results. (a) Fraction of responses (aggregated first within each user, then across users) for each option (dislikes, likes and superlikes) using Random recommender. (b) Mean user utility over time using LibFM recommender.
  We find that Treatment c has the highest level of utility for users. 
  }
 	\label{fig:sim_ctld_res}
 \end{figure}

  \begin{figure}[bt]
  \vspace{-0.2cm}
 	\centering
\subfloat[][Random]{
 		\includegraphics[width=.48\textwidth]{graphs/sim/exp_sim/Mean_user_ratio_for_rating_random.pdf}
    \label{fig:mean_rating_sim_exp_random}
}
 	\hfill
\subfloat[][TopPop]{
 		\includegraphics[width=.48\textwidth]{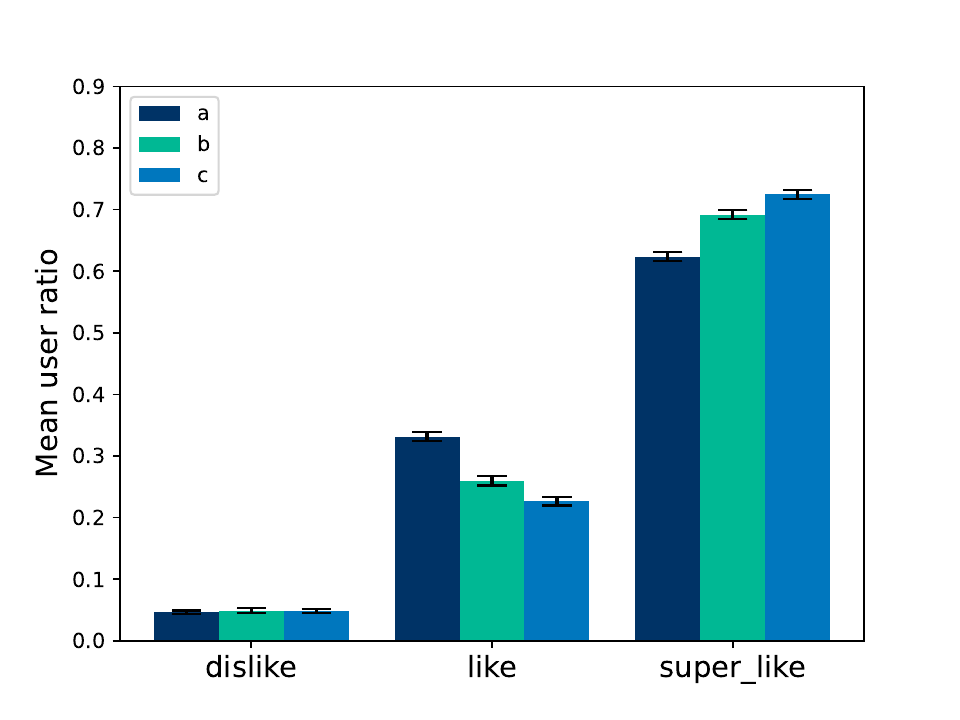}
    \label{fig:mean_rating_sim_exp_toppop}
}
 	\caption{Fraction of responses (aggregated first within each user, then across users) for each option (dislikes, likes and superlikes) in the simulated experiment.}
 	\label{fig:sim_exp_ratings}
 \end{figure}

  \begin{figure}[bt]
  \vspace{-0.2cm}
 	\centering
\subfloat[][Random]{
 		\includegraphics[width=.48\textwidth]{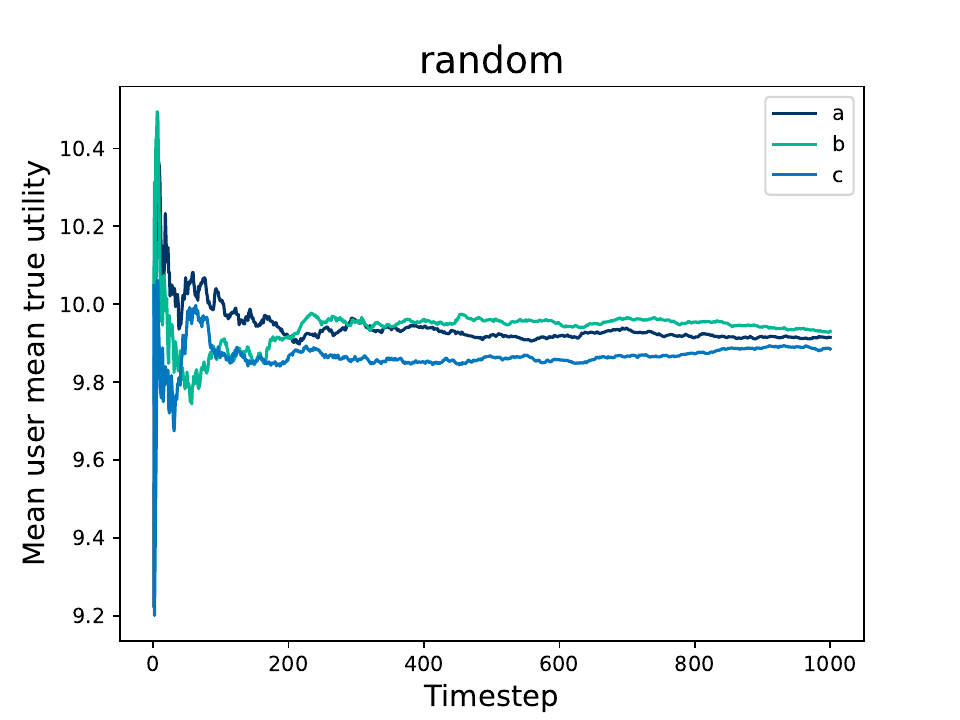}
    \label{fig:mean_utils_exp_sim_random}
}
 	\hfill
\subfloat[][TopPop]{
 		\includegraphics[width=.48\textwidth]{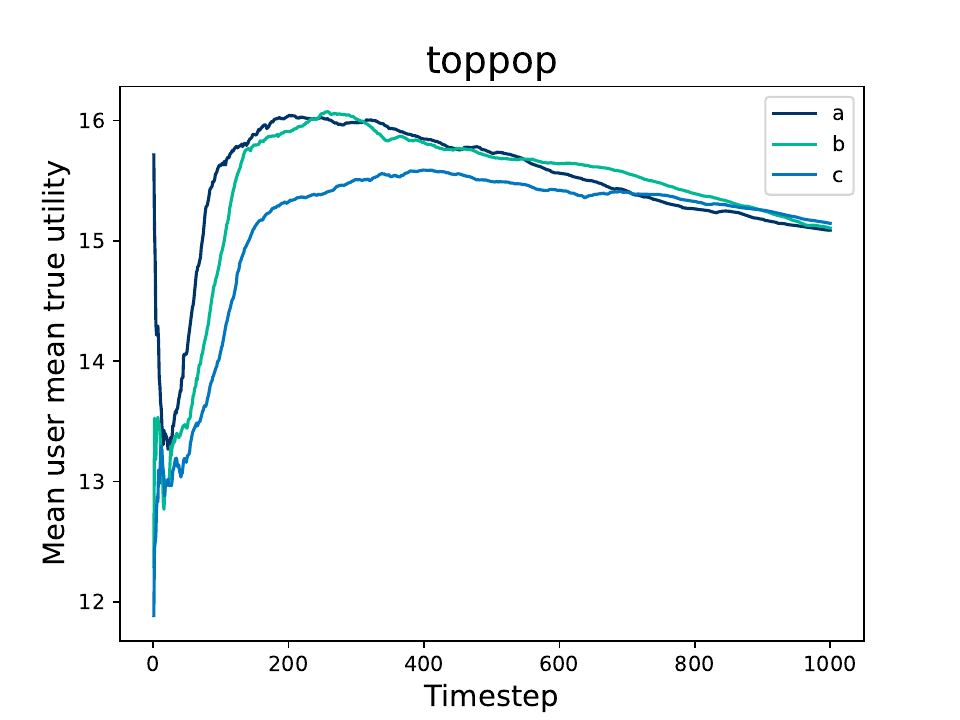}
    \label{fig:mean_utils_exp_sim_toppop}
}
 	\caption{Mean user utility over time in the controlled experiment using libFm and topPop recommender algorithms, for random recommender algorithm see Figure~\ref{fig:mean_rating_sim_exp_libfm}.}
 	\label{fig:sim_exp_utils}
 \end{figure}

  \begin{figure}[bt]
  \vspace{-0.2cm}
 	\centering
\subfloat[][LibFM]{
 		\includegraphics[width=.48\textwidth]{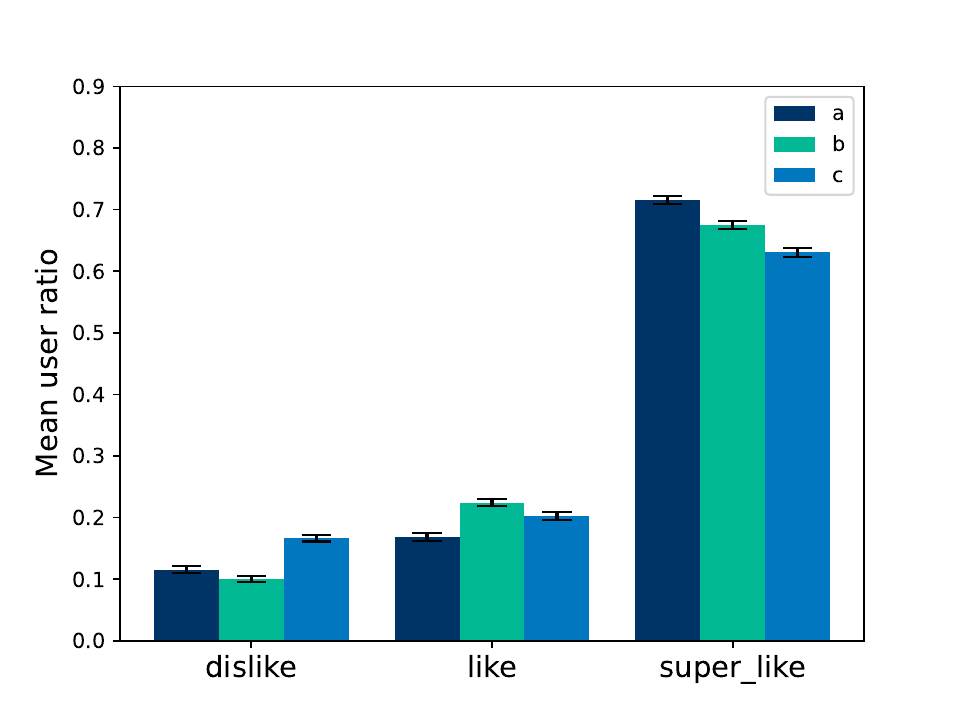}
    \label{fig:mean_rating_sim_libfm}
}
 	\hfill
\subfloat[][TopPop]{
 		\includegraphics[width=.48\textwidth]{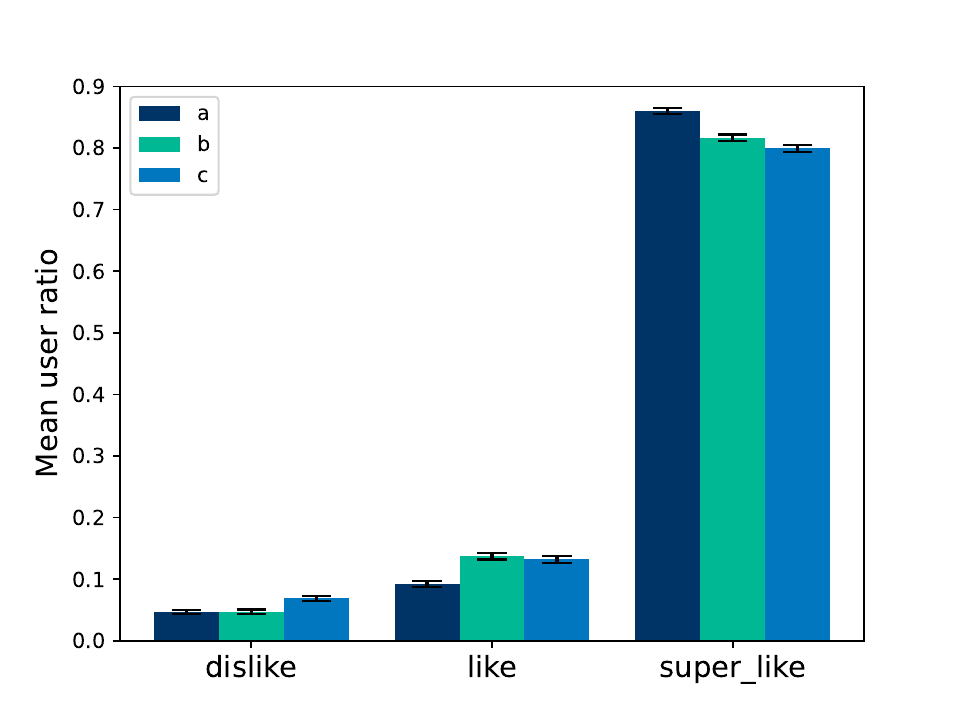}
    \label{fig:mean_rating_sim_toppop}
}
 	\caption{Fraction of responses (aggregated first within each user, then across users) for each option (dislikes, likes and superlikes) in the controlled simulation.}
 	\label{fig:sim_ratings}
 \end{figure}

  \begin{figure}[bt]
  \vspace{-0.2cm}
 	\centering

\subfloat[][LibFM]{
 		\includegraphics[width=.48\textwidth]{graphs/sim/controled_sim/Utility_over_time_libfm.pdf}
    \label{fig:mean_utils_sim_libfm}
}
 	\hfill
\subfloat[][TopPop]{
 		\includegraphics[width=.48\textwidth]{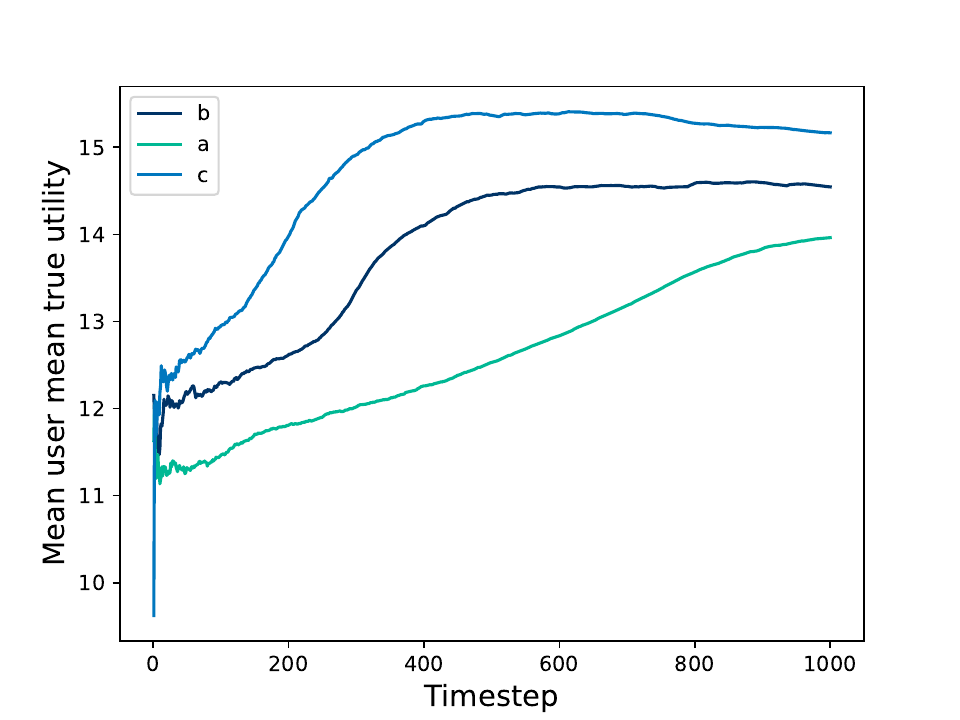}
    \label{fig:mean_utils_sim_toppop}
}
 	\caption{Mean user utility over time in the controlled experiment using libFm and topPop recommender algorithms, for random recommender algorithm see Figure~\ref{fig:mean_rating_sim_ctld_libfm}.}
 	\label{fig:sim_utils}
 \end{figure}

%% file: appendix_regressiontables.tex
 {\renewcommand\normalsize{\tiny}%
\normalsize
\begin{table}[tbh]
\input{plots_nikhil/experiment_likeregression.tex}
\caption{Analogue of \Cref{tab:natexpregression_explainratings}, but for the RCT. Regression results for a \textit{single} user-song rating (i.e., either a 0, 1, or 2) versus the average other ratings by that user or for that song. As before, this rating is more strongly associated by the \textit{user's} average behavior than the \textit{song's} average rating. Treatment c is slightly (but statistically significantly with $p < .05$) more informative in terms of the coefficient on the mean song rating by other users. On average, for each treatment group, each song has about 23 ratings; users have on average 37. Qualitatively similar results emerge for other filtering techniques.}
\label{tab:experiment_regression_explainratings}
\end{table}
}

 {\renewcommand\normalsize{\tiny}%
\normalsize
\begin{table}[tbh]
\input{plots_nikhil/experiment_meanregression_withpersonalized.tex}
\caption{Analogue of \Cref{tab:natexpregression_multiple_ratings}  for the RCT. Regression results for a song's mean rating score on a \textit{test} set versus \textit{train set} mean rating scores given by users (who gave the song test-set ratings) or for that song. The mean-song-rating coefficient is approximately the same as the mean-user-rating coefficient, suggesting that both song quality and user behavior contribute to a song's mean rating. Furthermore, we add a term for the fraction of times in the test set that a song appeared via \textit{personalized} recommendations; this fraction has a strong positive association with the average song rating in the test set, even controlling for song quality as determined in the training set. We find no significant differences between treatment groups. All independent and dependent variables are standardized. On average, in each treatment group, each song has about 12 ratings on average in each of the test and train set; users have on average 23 ratings in the train set. Qualitatively similar results emerge for other filtering techniques.}
\label{tab:experiment_regression_explainratings_multiple_withpersonal}
\end{table}
}

%% file: plots_nikhil/experiment_likeregression.tex
\begin{center}
\begin{tabular}{lclc}
\toprule
\textbf{Dep. Variable:}                                  & user\_song\_rating & \textbf{  R-squared:         } &     0.210   \\
\textbf{Model:}                                          &        OLS         & \textbf{  Adj. R-squared:    } &     0.209   \\
\textbf{Method:}                                         &   Least Squares    & \textbf{  F-statistic:       } &     686.9   \\
\textbf{No. Observations:}                               &        33701       & \\ 
\bottomrule
\end{tabular}
\begin{tabular}{lcccccc}
                                                         & \textbf{coef} & \textbf{std err} & \textbf{t} & \textbf{P$> |$t$|$} & \textbf{[0.025} & \textbf{0.975]}  \\
\midrule
\textbf{treatment[a]}                                    &       0.6599  &        0.006     &   107.139  &         0.000        &        0.648    &        0.672     \\
\textbf{treatment[b]}                                    &       0.6776  &        0.007     &   100.076  &         0.000        &        0.664    &        0.691     \\
\textbf{treatment[c]}                                    &       0.6573  &        0.006     &   105.885  &         0.000        &        0.645    &        0.669     \\
\textbf{personalized}                                    &       0.1303  &        0.008     &    15.910  &         0.000        &        0.114    &        0.146     \\
\textbf{mean\_user\_rating\_others}                      &       0.2590  &        0.006     &    45.910  &         0.000        &        0.248    &        0.270     \\
\textbf{mean\_user\_rating\_others:treatment[T.b]}       &       0.0177  &        0.009     &     2.068  &         0.039        &        0.001    &        0.034     \\
\textbf{mean\_user\_rating\_others:treatment[T.c]}       &       0.0144  &        0.008     &     1.785  &         0.074        &       -0.001    &        0.030     \\
\textbf{mean\_song\_rating\_others}                      &       0.1320  &        0.006     &    22.858  &         0.000        &        0.121    &        0.143     \\
\textbf{mean\_song\_rating\_others:treatment[T.b]}       &       0.0169  &        0.009     &     1.977  &         0.048        &        0.000    &        0.034     \\
\textbf{mean\_song\_rating\_others:treatment[T.c]}       &       0.0356  &        0.008     &     4.414  &         0.000        &        0.020    &        0.051     \\
\textbf{user\_ratings\_count}                            &       0.0035  &        0.003     &     1.018  &         0.309        &       -0.003    &        0.010     \\
\textbf{mean\_user\_rating\_others:user\_ratings\_count} &       0.0196  &        0.004     &     5.491  &         0.000        &        0.013    &        0.027     \\
\textbf{song\_ratings\_count}                            &       0.0297  &        0.004     &     8.008  &         0.000        &        0.022    &        0.037     \\
\textbf{mean\_song\_rating\_others:song\_ratings\_count} &       0.0299  &        0.003     &     8.590  &         0.000        &        0.023    &        0.037     \\
\bottomrule
\end{tabular}
\end{center}


%% file: plots_nikhil/experiment_meanregression_withpersonalized.tex
\begin{center}
\begin{tabular}{lclc}
\toprule
\textbf{Dep. Variable:}                                        & mean\_song\_rating\_test & \textbf{  R-squared:         } &     0.340   \\
\textbf{Model:}                                                &           OLS            & \textbf{  Adj. R-squared:    } &     0.334   \\
\textbf{Method:}                                               &      Least Squares       & \textbf{  F-statistic:       } &     57.38   \\
\textbf{No. Observations:}                                     &            1464          & \\
\bottomrule
\end{tabular}
\begin{tabular}{lcccccc}
                                                               & \textbf{coef} & \textbf{std err} & \textbf{t} & \textbf{P$> |$t$|$} & \textbf{[0.025} & \textbf{0.975]}  \\
\midrule
\textbf{treatment[a]}                                          &       0.0034  &        0.035     &     0.098  &         0.922        &       -0.066    &        0.073     \\
\textbf{treatment[b]}                                          &      -0.0282  &        0.040     &    -0.707  &         0.480        &       -0.106    &        0.050     \\
\textbf{treatment[c]}                                          &      -0.0007  &        0.037     &    -0.018  &         0.985        &       -0.073    &        0.071     \\
\textbf{frac\_personalized\_test}                              &       0.1819  &        0.024     &     7.455  &         0.000        &        0.134    &        0.230     \\
\textbf{mean\_user\_rating\_train}                             &       0.3485  &        0.036     &     9.688  &         0.000        &        0.278    &        0.419     \\
\textbf{mean\_user\_rating\_train:treatment[T.b]}              &      -0.0228  &        0.053     &    -0.427  &         0.669        &       -0.127    &        0.082     \\
\textbf{mean\_user\_rating\_train:treatment[T.c]}              &      -0.0182  &        0.051     &    -0.356  &         0.722        &       -0.118    &        0.082     \\
\textbf{mean\_song\_rating\_train}                             &       0.3668  &        0.037     &     9.997  &         0.000        &        0.295    &        0.439     \\
\textbf{mean\_song\_rating\_train:treatment[T.b]}              &       0.0568  &        0.053     &     1.064  &         0.288        &       -0.048    &        0.162     \\
\textbf{mean\_song\_rating\_train:treatment[T.c]}              &      -0.0024  &        0.051     &    -0.046  &         0.963        &       -0.102    &        0.098     \\
\textbf{count\_song\_ratings\_train}                           &       0.1411  &        0.026     &     5.457  &         0.000        &        0.090    &        0.192     \\
\textbf{mean\_song\_rating\_train:count\_song\_ratings\_train} &       0.1313  &        0.025     &     5.274  &         0.000        &        0.082    &        0.180     \\
\textbf{count\_user\_ratings\_train}                           &      -0.0450  &        0.024     &    -1.904  &         0.057        &       -0.091    &        0.001     \\
\textbf{mean\_user\_rating\_train:count\_user\_ratings\_train} &       0.0184  &        0.021     &     0.870  &         0.385        &       -0.023    &        0.060     \\
\bottomrule
\end{tabular}
\end{center}


%% file: plots_nikhil/experiment_uservariance.tex
\begin{tabular}{llll}
\toprule
{} &              All &           Random &     Personalized \\
\midrule
a &  0.0924 ± 0.0131 &  0.1012 ± 0.0147 &  0.1231 ± 0.0176 \\
b &  0.1066 ± 0.0186 &  0.1215 ± 0.0203 &  0.1447 ± 0.0226 \\
c &  0.1205 ± 0.0193 &  0.1145 ± 0.0176 &  0.1606 ± 0.0279 \\
\bottomrule
\end{tabular}

%% file: plots_nikhil/experiment_meanregression.tex
\begin{center}
\begin{tabular}{lclc}
\toprule
\textbf{Dep. Variable:}                                        & mean\_song\_rating\_test & \textbf{  R-squared:         } &     0.314   \\
\textbf{Model:}                                                &           OLS            & \textbf{  Adj. R-squared:    } &     0.309   \\
\textbf{Method:}                                               &      Least Squares       & \textbf{  F-statistic:       } &     55.44   \\
\textbf{No. Observations:}                                     &            1464          & \\
\bottomrule
\end{tabular}
\begin{tabular}{lcccccc}
                                                               & \textbf{coef} & \textbf{std err} & \textbf{t} & \textbf{P$> |$t$|$} & \textbf{[0.025} & \textbf{0.975]}  \\
\midrule
\textbf{treatment[a]}                                          &       0.0039  &        0.036     &     0.107  &         0.915        &       -0.067    &        0.074     \\
\textbf{treatment[b]}                                          &      -0.0298  &        0.041     &    -0.734  &         0.463        &       -0.109    &        0.050     \\
\textbf{treatment[c]}                                          &       0.0015  &        0.037     &     0.040  &         0.968        &       -0.072    &        0.075     \\
\textbf{mean\_user\_rating\_train}                             &       0.3549  &        0.037     &     9.688  &         0.000        &        0.283    &        0.427     \\
\textbf{mean\_user\_rating\_train:treatment[T.b]}              &      -0.0043  &        0.054     &    -0.079  &         0.937        &       -0.111    &        0.102     \\
\textbf{mean\_user\_rating\_train:treatment[T.c]}              &      -0.0119  &        0.052     &    -0.229  &         0.819        &       -0.114    &        0.090     \\
\textbf{mean\_song\_rating\_train}                             &       0.4088  &        0.037     &    11.068  &         0.000        &        0.336    &        0.481     \\
\textbf{mean\_song\_rating\_train:treatment[T.b]}              &       0.0518  &        0.054     &     0.952  &         0.341        &       -0.055    &        0.159     \\
\textbf{mean\_song\_rating\_train:treatment[T.c]}              &      -0.0074  &        0.052     &    -0.141  &         0.887        &       -0.109    &        0.095     \\
\textbf{count\_song\_ratings\_train}                           &       0.0662  &        0.024     &     2.729  &         0.006        &        0.019    &        0.114     \\
\textbf{mean\_song\_rating\_train:count\_song\_ratings\_train} &       0.1364  &        0.025     &     5.380  &         0.000        &        0.087    &        0.186     \\
\textbf{count\_user\_ratings\_train}                           &      -0.0558  &        0.024     &    -2.325  &         0.020        &       -0.103    &       -0.009     \\
\textbf{mean\_user\_rating\_train:count\_user\_ratings\_train} &       0.0276  &        0.022     &     1.278  &         0.201        &       -0.015    &        0.070     \\
\bottomrule
\end{tabular}
\end{center}
